\documentclass{elsarticle}
\usepackage{graphicx} 
\usepackage{url} 
\usepackage{xcolor} 

\usepackage{listings}
\lstdefinelanguage{Kotlin}{
    keywords={data, class, val, var, fun, return, if, else, for, in, when},
    morecomment=[l]{//},
    morestring=[b]"
}
\lstdefinelanguage{json}{
    basicstyle=\ttfamily\small,
    string=[s]{"}{"},
    stringstyle=\color{blue},
    morecomment=[l]{//},
    commentstyle=\color{gray},
    showstringspaces=false
}

\usepackage{algorithm}
\usepackage{algpseudocode}
\usepackage{amsmath} 
\algdef{SE}[SWITCH]{Switch}{EndSwitch}[1]{\textbf{switch} #1}{\textbf{end switch}}
\algdef{SE}[CASE]{Case}{EndCase}[1]{\textbf{case} #1}{\textbf{end case}}
\algtext*{EndSwitch}
\algtext*{EndCase}

\usepackage{soul}

\begin{document}
\title{A Real-time tax compliance architecture for VAT split payments with Making Tax Smart and Open Banking.}
\author{Panos Louvieris, George Ioannou, Gareth White}
\affiliation{Department of Computer Science, Brunel University of London, Uxbridge, UK}
\date{August 2026}

\begin{abstract}
This paper presents Making Tax Smart with Open Banking (MTS-OB), a novel DLT-based architecture designed to integrate real-time VAT compliance into the existing payment system through split payments and Open Banking. Motivated by the substantial VAT gap, the research investigates how leveraging the informational value embedded within payment transactions can reduce the tax administration burden. MTS-OB, a proof-of-concept application built on R3 Corda, facilitates automated, real-time VAT split payments (VSP) without disrupting current payment infrastructure. MTS-OB integrates with existing banking systems via Open Banking APIs, enabling secure information management for VAT collection, reporting, analytics, and automated compliance enforcement. The system uses traditional bank deposit money as a real-time information sensor thereby extending its functionality to include "money as a medium of real-time taxation". Scalability testing demonstrates the system's ability to maintain operational stability under increasing transaction loads, illustrating its potential for real-world deployment. The key finding is the necessity of redefining money to include its informational value for real-time tax compliance, requiring a redesign of the money system while adhering to existing regulations and standards. This research demonstrates the feasibility of integrating tax functionality directly into payments, offering a path towards "hard to get tax wrong" compliance and a potential reduction in the VAT gap without compromising financial stability.
\end{abstract}

\maketitle

\section{Introduction}
The difference between the tax legally owed by taxpayers and the tax which is actually paid is called the “tax gap”. Part of the tax gap is caused by inadvertent errors in tax statements; however, the biggest proportion is due to tax avoidance, evasion and fraud. It is estimated that in the UK alone vital public services are deprived of roughly £46.8 billion of revenue yearly because of the tax gap \cite{hmrc2025taxgaps}. This gap is associated with several taxes such as Corporation Tax, Income Tax, National Contributions, Capital Gains Tax, Excise Duties and Value Added Tax, commonly referred to as VAT. The research presented herein focuses on VAT which is estimated at 6.6\% of the theoretical VAT liability and amounts to £12.1 billion for the 2024-2025 fiscal year \footnote{https://www.gov.uk/government/statistics/measuring-tax-gaps/2-tax-gaps-vat}.
VAT is a tax paid on sales, therefore every trade or service transaction incurs tax which amounts to 20\%, 5\% or 0\% of the net value. The current regime, where the supplier of goods or services acts as the tax collection agent on behalf of HMRC, was established in an era where payment and accounting systems were offline and paper-based hence unreliable and slow. Furthermore, the VAT reporting processes are prone to error and fraud because responsibility for the validity of information is left with the taxpayer. Specifically in the UK, His Majesty’s Revenue \& Customs (HMRC), the UK tax authority, has taken steps towards digitisation with Making Tax Digital (MTD) \cite{valenta2017comparison} ; however, real-time processing of tax information and VAT payments are still not applicable and suitable for a science and technology savvy 21st century digital economy. A tax administration reform is necessary, where taxes are calculated, collected and reported based on timely and accurate information which is generated, recorded and validated at the point of the transaction. The research presented in this paper has been conducted from a UK perspective in observing payments, tax, law, regulation and standards.

This information is readily available, yet unutilised for tax purposes, through modern Point-of-Sale systems, such as those found on the high street and those present in online shops. These POS systems have been developed and maintained for the past decades by banks and affiliated payment system vendors with the highest levels of reliability and security, in accordance with industry standards. Examples of standards include; the ISO 20022 which describes the data structures that facilitate payment information exchange, and Open Banking which enables financial institutions to share structured payments data securely through Application Programming Interfaces (APIs). In response to payments innovations facilitated by the banking sector, citizens and businesses have gradually replaced banknotes (which do not require a POS) with credit cards and mobile payment apps that interface with bank deposits. 
It follows that a tax administration reform based on timely and accurate information should be underpinned by the established payment ecosystem where bank deposits are integral. This reform would not have been possible without the emergence of the Open Banking APIs and the availability of payments information which is enforced by law. Open Banking has opened access to hitherto siloed payments data and has provided control to data subjects i.e. consumers and businesses who make payments, over their transaction information. More importantly, access to payments can be exercised in real-time, a key requirement for tax administration reform. 

It is therefore understood that the key information requirements to achieve a new capability for tax administration such as timeliness, availability and accuracy, are fulfilled with existing banking and payment standards and technology. To enable new real-time tax payments and assure the privacy of taxpayers, it is also necessary to develop the infrastructure capability for managing the information variety which has now become available. Existing payment systems provide the necessary functional elements for tax administration reform however lack in transparency concerned with taxpayers’ data processing, which we perceive as the bedrock of trust between the tax authority and the citizen. Therefore, a technological framework that provides visibility and accountability of information processing is essential for the success of the tax administration reform. Distributed Ledger Technology, and in particular the R3 Corda DLT are deemed appropriate for this task.

The R3 Corda is a permissioned and private distributed ledger technology platform, specially designed for financial industry applications and with smart contract support \cite{valenta2017comparison}. In contrast to other DLT frameworks, transactions in Corda are only announced to relevant parties, with all communication in a Corda network taking the form of small multiparty protocols, called flows \cite{khan2017distributed}. For the first time, an approach takes advantage of the unique permission and access-control management of Corda to make payments trilateral instead of bilateral transactions, causing money to act in a different, novel role is proposed. Moreover, Corda has been designed by the R3 consortium which consists of private banks and has been the DLT of choice for government applications, such as Central Bank Digital Currencies.

The points made so far, form part of the specification for a tax innovation which is more than a digitisation of the current paper-based and asynchronous tax system but instead is built into the money system to augment payments with tax functionality and extend the utility and definition of money. Tax collection has been integral to economic prosperity, however it is still viewed as an inhibitor for innovation, partly due to the administrative overhead it imposes on governments and businesses alike. This overhead is explained by the lack of connectivity between two conceptual ledgers. The ledger which holds money transfers and the ledger which holds tax information. These ledgers have been insofar, asynchronous. For the first time, the two ledgers become connected and synchronous, through an infrastructure capability which is invisible to the end user and without the need for introducing any new legislation. 

The purpose of the research presented herein is to challenge the existing money and tax systems and develop the infrastructure capability which can be built upon the existing money system to effectively extend its functionality and achieve a new capability for tax administration. We present Making Tax Smart with Open Banking (MTS-OB), an infrastructure capability for the HMRC’s reform of tax administration in accordance with the HMRC Transformation Roadmap \cite{hmrctransformation}. This is an extension of previous research concerned with automated tax collection with a Central Bank Digital Currency \cite{louvieris2024making}. For the first time, a DLT-enabled tax administration architecture which enables VAT split payments (VSP) to be made in real-time through Open Banking using the most trusted form of money available today to businesses and citizens i.e. bank deposit money, is proposed and demonstrated.

Although MTS-OB is an invisible layer between third-party providers and Open Banking (see section \ref{mtsoboverview}), it operates on the payment path rather than alongside it, which causes every VAT-liable transaction to pass through it. Any architecture proposed for this role must therefore behave predictably as transaction volumes and participant numbers grow. An approach which is functionally correct but degrades unpredictably under load would be unusable regardless of its benefits for tax compliance. 

Establishing scalability of the proposed architecture, i.e. how it responds to increasing load, is therefore a necessary condition for confirming feasibility, as shown in this paper.

\subsection{Aim}
The aim of the research presented in this paper is to design and evaluate a permissioned distributed ledger architecture which demonstrates the feasibility of extending the functionality of money to support automated tax payments using bank deposit money.

\subsection{Objectives}
\begin{enumerate}
    \item Identify the challenges and constraints (regulatory, technical)  for the design of a modern tax system that supports HMRC in leveraging the hidden informational value and intelligence within payments for tax administration purposes.
    \item Develop a proof-of-concept Open Banking compliant ‘Making Tax Smart’ application, which does not disrupt the existing payment system but enables real-time automated tax compliance.
    \item Demonstrate the feasibility and scalability of the developed ‘Making Tax Smart’ application for smart contract-enabled VAT split payments.
\end{enumerate}

The remainder of this paper is structured as follows: Section 2 establishes the background by examining the persistent challenge of tackling the VAT gap, the role of money in history and its evolution in recent years, the impact of the money system on tax administration and the international standards that facilitate safe transition to new money and tax system. Section 3 details the architecture, infrastructure capabilities, and technical specification of the Making Tax Smart with Open Banking (MTS-OB) prototype, including its integration with R3 Corda, smart contracts, states, flows, and Open Banking APIs. Section 4 presents the performance factors, testbed and prototype used to evaluate the scalability of MTS-OB. Section 5 presents the experimental results assessing transaction throughput and latency under varying workloads, transaction loads, and client counts. Section 6 discusses the broader implications of the findings in the context of digital economy transformation and monetary policy. Section 7 highlights ongoing future work within the ATTACC project, and Section 8 concludes with a summary of the contributions made by the framework.

\section{Background}
\subsection{The VAT Gap}
VAT fraud, evasion and avoidance are sources of money haemorrhaging for tax administrations which put pressure on government spending and creates injustices for law abiding taxpayers and market competition. In the UK alone, it is estimated at 6.6\% of the theoretical VAT liability and amounts to £12.1 billion for the 2024-2025 fiscal year. The gap, which is estimated using national accounts data and actual cash receipts reported by taxpayers, is the difference between the tax legally owed by taxpayers and the tax which is actually paid. It is caused by tax fraud, tax evasion, tax avoidance and optimisation practices, bankruptcies, financial insolvencies, as well as administrative errors and thus remains an important problem at a time when “governments need sustainable revenues to help weather today’s economic uncertainty”. In particular, the UK has increased its efforts in reducing the tax gap, focusing on making it “much harder for people to avoid paying the tax they owe” \cite{hmrc2025taxgaps}.

Moreover, VAT is a tax imposed on transactions that involve the exchange of goods and services, which inform key economic variables for government, such as growth, inflation, balance of payments, productivity and more. Therefore, the associated transaction data carry immense information value for economic policies i.e. monetary, financial and fiscal, further amplifying the value of VAT for the economy. As a response, the Office for National Statistics (ONS), has recently made use of data provided in VAT returns to track growth in close-to-real-time and understand the behaviour of individual industry sectors.

\subsection{Evolution of Money}

Money is central to the functioning of the UK economy, its tax administration system and the Government’s delivery of fiscal policy for economic and societal benefit. Despite the fact we live in a predominantly digital economy with the ever increasing use of digital fiat money (i.e. bank deposit money held in online bank accounts) when making payments for goods and services, the functions of money have remained unchanged from the earliest times e.g. circa 630 BC when the Lydians invented regulated coinage money \cite{melitz2017model}. In 1875 William Jevons first analysed and listed the functions of money as (i) a medium of exchange, (ii) a common measure of value (i.e. a unit of account), (iii) a store of value, and (iv) a standard of value (i.e. a standard of deferred payment) \cite{milnes1919economic}; where, (iv) is not an essential function of money. Moreover, the Bank of England, in their March 2020 CBDC discussion paper, still adheres to the functions of money as “… serving as (a) a store of value with which to transfer purchasing power from today to the future; (b) a medium of exchange with which to make payments for goods and services, and (c) a unit of account with which to measure the value of a particular good, service, savings product or loan.” \cite{boe2020cbdc}.

\subsection{Cryptoassets, Stablecoins and the need for Trust}\label{crypto}
Other forms of pseudo “digital money” are cryptocurrencies (also known as cryptoassets) such as Bitcoin and Ethereum’s Ether, which are not backed by a trusted central authority. They are not digital money because they do not perform the essential functions of money as set out above and are too volatile to be a reliable store of value. What is more, they are not widely accepted as a means of exchange and neither are they used as a unit of account \cite{bankofengland_what_is_money} . For instance, during May 2022, notably the TerraUSD stablecoin that was pegged to the US dollar catastrophically collapsed . Therefore, stablecoins cannot be considered to be a viable and reliable form of digital money either; they are another unbacked cryptoasset, unregulated and operate outside the supervision of central banks \cite{lords2022cbdc}.

So far in this discussion the evidence presented above demonstrates that all the above mentioned blockchain-based digital money types do not conform to or carry out all the necessary functions of money that would be mandatory and necessary to support any innovation for the development of a modern tax system. In today’s digital economy, bank deposits are the default type of digital money that is trusted by everyone in the economy and used as the medium of exchange by businesses and consumers alike for most payment transactions \cite{mcleay2014money}. The evidence for this is the majority of money (96\%) is held electronically as bank deposits with the remaining much smaller proportion (4\%) kept as physical cash \cite{mcleay2014money}. Bank deposits are the digital currency and one of the foundation stones on which the reform of tax administration must be built i.e. trust. Therefore, the design of Making Tax Smart, the infrastructure capability for UK’s tax administration reform which is presented in this paper, is designed around the concept of maintaining trust.

\subsection{The Impact of Central Bank Digital Currencies on Tax}
The Central Bank Digital Currency (CBDC) is an alternative form of digital money which is being considered by central banks around the world; the e-krona by Riksbank \cite{riksbank2021ekrona}, Project Rosalind \cite{bis2022rosalind}, Project mBridge \cite{bis2022mbridge} and the Digital Pound by the Bank of England \cite{boe2024digitalpound}, to name a few. We consider the consultation issued by the BoE for this paper, since the use case is VAT compliance for the purposes of HMRC.	

The UK’s design for CBDC, called the Digital Pound (DP), involves the distribution of digital cash via private intermediaries, called Payment Service Providers (PSPs). DPs and wallet balances are kept in a central ledger which is maintained by the BoE, and the PSPs hold pass-through wallets of end users which provide access to the ledger via an API. The BoE’s blueprint for a DP, states: “the DP could improve resilience as an additional payment rail” \cite{boe2023consultation} hence would not disrupt or interfere with standard payment rails that rely on bank deposits. This key design aspect, in addition to the use of APIs means that the envisioned Making Tax Smart platform presented in this paper does not clash with a future implementation of a DP; instead, it will be readily interoperable with DP via APIs, similarly to how bank deposits are accessible via Open Banking APIs. Whilst the BoE has not yet made a decision on whether it will introduce a DP \cite{boe2025update}, the DP blueprint remains compatible with the ‘Making Tax Smart with Open Banking’ approach proposed in this paper. 

\subsection{Legal and Regulatory Compliance Aspects}
Compliance with the existing regulatory frameworks and adherence to industry standards is an additional key design requirement of Making Tax Smart with Open Banking. This is an essential element that ensures faster implementation and streamlined integration with the existing regulatory landscape in payments and banking, including the recent standardisation efforts in the digitisation of trade documents. 

The Open Banking Standard which is overseen by Open Banking Limited (OBL) formerly known as Open Banking Implementation Entity (OBIE) \cite{obie2024standards} prescribes the rules of Open Banking services with reference to the Application Programming Interfaces (APIs) that banks must freely provide to Third-party Providers (TPPs) and the security procedures that must be followed. Through Open Banking, end-users can grant Account Information Services (AIS) and Payment Initiation Services (PIS) permissions to TPP’s. Enhancing competition via Open Banking in payments \cite{buckley2020road} creates a drive for transformation of finance through technology \cite{dratva2020open} enabling third parties to innovate and create new financial products \cite{lords2022cbdcb} and services. The benefits of Open Banking, equally apply to government. HMRC can make use of Open Banking for tax administration purposes, for example to streamline tax compliance with automated tax payments.

Further, standardisation efforts in the digitisation of trade documents are under way, covering electronic trade documents \cite{uncitral2017mletr, etda2023}, and e-invoicing via the Peppol network which will become mandatory in the UK in April 2029 \cite{hmrc2026taxupdate}). Moving away from traditional paper-based transaction to digital files in structured format offers many benefits such as, improved auditing processes, improved traceability of orders, rapid access and retrieval \cite{hmrc2024einvoicing}. This paradigm shift becomes important in light of the ISO20022 standard adoption \cite{swift2020iso, hmtreasury2024nationalpayments}. ISO20022 introduces structured fields within payment messages which can carry contextual information and metadata associated with the payment, such as invoices. Invoices which follow the EN16931 standard \cite{ec2026en16931compliance} contain information which is used for tax purposes i.e. calculating VAT payments. Therefore, consolidating payment information with the invoices will dramatically increase the availability of transaction data which contains economic intelligence of immense value that government departments can use for policy making and risk management. HMRC already has the right to inspect that financial data under Finance Act 2008 Schedule 36 \cite{financeact2008}.

It follows that Open Banking, the ISO20022 standard and the digitisation of trade documents provide the sufficient and necessary regulatory and standards backdrop for automated VAT compliance using bank deposit money. Therefore the transformation of the money system should not be premised on DLT-enabled digital currencies e.g. cryptoassets and stablecoins, which carry disproportionate or even unknown risk profiles as discussed in section \ref{crypto}. Instead, the focus should be on making the best of new technologies and on maximising the gains from existing regulations around payments and information sharing. 

\subsection{Extending the Functionality of Money: Achieve new Functionality of Money}

In today’s data-driven digital economy the three current functions of money (defined above) are not sufficient because they do not reflect the Informational Value of money that would be required for tax administration purposes relating to the real-time payment and reporting of VAT to HMRC in order to address the annual VAT gap. 

The concept of Smart Money was originally introduced in \cite{louvieris2024making}for extracting the informational value of money from CBDC's using Corda Distributed Ledger technology and smart contracts. In this paper, for the first time, the Smart Money concept has been developed and applied to bank deposit money for the purpose of extracting the informational value from payments hence extending the functionality of money to include “money as a real-time information sensor”.

The proof-of-concept implementation of Smart Money for real-time split payment of VAT directly to HMRC’s bank account is the ‘Making Tax Smart with Open Banking’ VAT CorDapp application (MTS-OB for short). It is important to note that MTS-OB does not interfere with or disrupt existing payment rails but acts as an invisible layer between third-party providers and Open Banking. 

R3 Corda is a permissioned and private distributed ledger technology platform specially designed for financial industry applications and with smart contract support. In contrast to other DLT frameworks, transactions in Corda are only announced to the involved parties on a need-to-know basis. The unique permission and access–control management adopted in Corda is aligned with MTS-OB approach in terms of secure data sharing, where payments become trilateral instead of bilateral transactions \cite{louvieris2024making}, so that the tax authority i.e. HMRC has the required visibility to assure the collection of taxes for the associated transactions. 

The best way to think of MTS-OB is as a cyber-secure precision information management architecture that facilitates the real-time collection of VAT, VAT reporting and analytics, and automated VAT compliance enforcement. Thus, making it ‘hard to get tax wrong’ for businesses \cite{hmrc2025taxgaps}. Moreover, MTS-OB offers the real prospect of unobtrusively building tax functionality into money (i.e. a medium of real-time taxation) without causing financial instability.

\section{MTS-Open Banking Prototype}

\subsection{Why go the Open Banking route?}
Open Banking is the practice of making bank customer information available to TPPs, following permission provided by the customers themselves. Bank customer information is provided by the banks via APIs, so that this information becomes available almost instantly and can therefore be used for payment and information services by the TPP. The banks are responsible for supplying the payments information (e.g. bank account, account balance) to the TPP in a secure manner to protect their customers’ rights to privacy. With most of the retail money being held in bank deposits, banks are a critical part of the payments ecosystem since they hold the funds that make these payments real. Therefore, any major innovation and transformation of the money system must consider banks as a critical component of the new landscape and must make the best of the services that they offer. Further, Open Banking has successfully been integrated into the payments system, it has given rise to multiple new services such as account management, loans and insurance, it has demonstrated resilience and has been established as a component that ensures customer privacy and control of personal data. The regulatory and supervisory frameworks for overseeing its use are already in place \cite{obie2024regulatory}. It follows that Open Banking is the ideal gateway to retail money and the suitable entry-point for any payment innovation that does not interfere with the money system and does not pose additional risks to financial stability.

The MTS-OB acts as an information management infrastructure capability for UK payments which enables a broad range of cross-cutting smart data sharing services for government departments; such as, automated tax compliance (HMRC), social benefits (DWP), criminal investigations (HO), and monitoring of the economy (ONS). Smart Money, via the MTS-OB platform, is involved with every payment that takes place within and across the border for tax collection purposes such as VAT and customs duties. With the use of cash in decline, the vast majority of payments is made using money held in bank deposits. It follows that Open Banking is the path of least resistance for incorporating Smart Money in the payment system. This is justified by a number of reasons:

\begin{enumerate}
    \item Interfacing with a bank account via Open Banking APIs is straightforward and tested.
    \item No new risks to financial stability are introduced by integrating Smart Money with Open Banking.
    \item Open Banking APIs are constantly being updated to integrate new requirements i.e. Variable Recurring Payments, Green Financing, e-invoicing.
    \item The Open banking route does not challenge the traditional role of banks as bank account holders and consequently, as lending institutions.
\end{enumerate}
\subsection{Open Banking Sandboxes}
The financial institutions that are obligated to provide open banking services also provide test instances for their open banking APIs. These instances are called sandboxes and their purpose is to facilitate TPPs to test their products before they are rolled out to their customers. If a TPP application is tested successfully on a sandbox, it means that it can interact with the corresponding production environment flawlessly. UK banks such as HSBC \cite{hsbc2024openbanking} and NatWest \cite{natwest2024sandbox} provide access to their open banking API sandboxes to registered TPPs. The Open Bank Project \cite{obp2024}, a software company which creates open banking API solutions for banks, makes such APIs available to TPPs and non-TPPs for developing their own products. The Open Bank Project provides, as open source, APIs which are in accordance with the standards specified by the OBL therefore are compliant to the financial regulations.

\subsection{Integrating MTS with Open Banking APIs}
The MTS with Open Banking couples DLT-enabled VSPs using bank deposit money which become available via Open Banking APIs. This is achieved by submitting the API calls and collecting the corresponding responses from within the Corda flows that implement the payments, i.e. from within the MTS architecture. This permits the realisation of VSPs with real, central-bank backed money that are kept in bank deposits without any intervention or disruption of the core banking system. 
The current iteration of MTS presented herein utilises two Open Banking services: Account Information Services (AIS) and Payment Information Services (PIS). AIS enables MTS to query bank account balances in order to initiate a payment only if the required funds are available in the payer’s account. PIS enables MTS to transfer the funds among the counterparties namely the payer, the payee and the HMRC. A description of the architecture and the payment journey is presented in the remainder of this section.

\subsection{MTS-OB overview}
\label{mtsoboverview}
\begin{figure}
    \centering
    \includegraphics[width=1\linewidth]{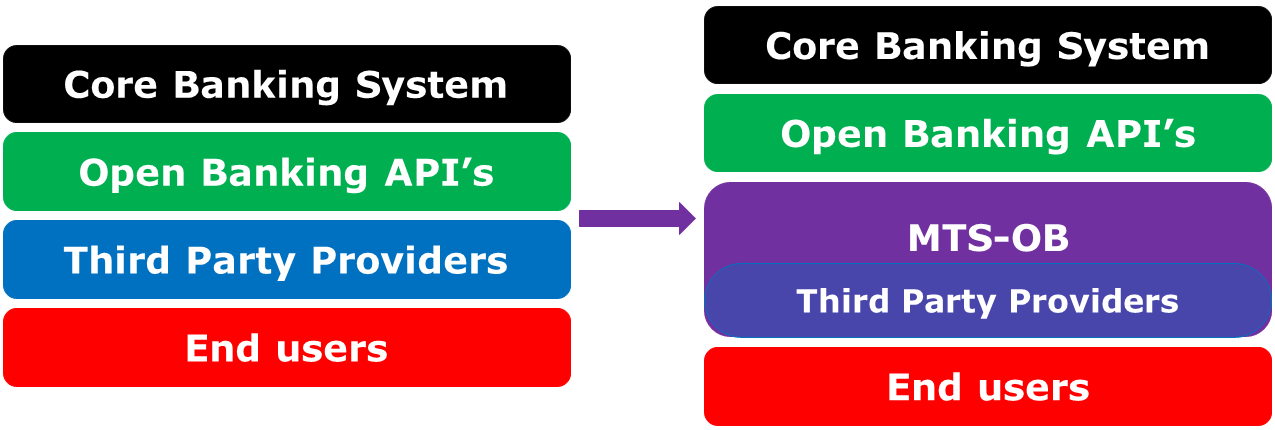}
    \caption{From Open Banking to MTS-Open Banking}
    \label{fig:openbankingtomtsob}
\end{figure}

\begin{figure}
    \centering
    \includegraphics[width=1\linewidth]{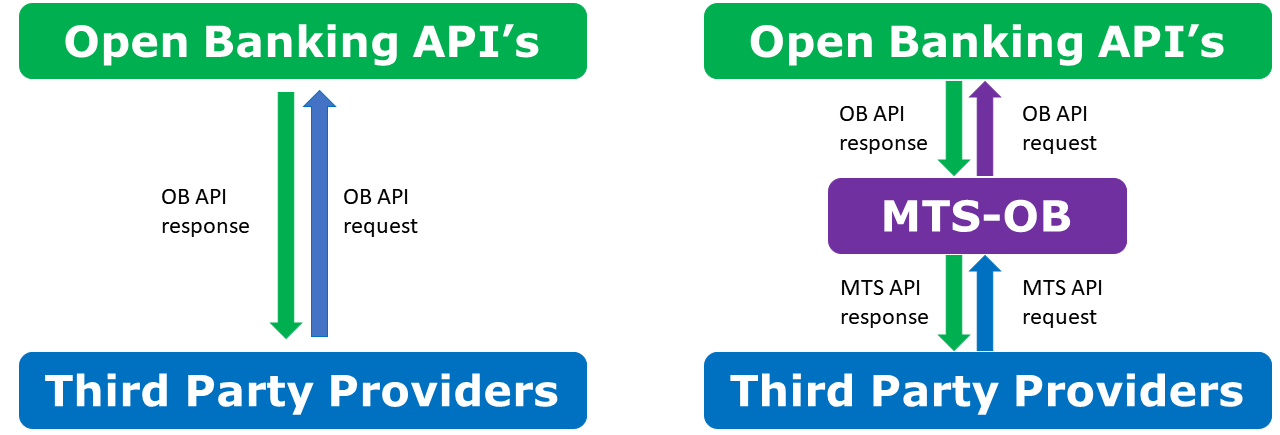}
    \caption{TPPs interaction with Open banking API's before MTS-OB (left) and after MTS-OB (right)}
    \label{fig:openbankingtomtsobinteraction}
\end{figure}

Making Tax Smart with Open Banking operates as an invisible layer between the end users (buyers and sellers) that are involved in transactions and the core-banking system. This layer is placed directly between TPPs and the Open Banking APIs presented by banks. With reference to Figure \ref{fig:openbankingtomtsob}, the building blocks of the new infrastructure capability are:

\begin{enumerate}
    \item Core Banking System: these are the back-end systems that process daily banking transactions. These systems, and the network thereof are maintained by banks.
    \item Open Banking APIs: these are the interfaces that banks present to TPPs in accordance with the Open Banking regulations. These APIs serve as the front-end of the bank accounts that are part of the Core Banking System.
    \item MTS-OB: this is the R3 Corda-based information management technology that enables a multitude of innovative services such as VSPs and programmable money. This layer is placed between Open Banking and End users, encompassing the TPP sub-layer.
    \item Third-party Providers: in accordance with Open Banking regulation, these are the organisations that use APIs made available by banks to access customers’ accounts. There are two categories, Payment Initiation Service Providers (PISPs) and Account Information Service Providers (AISPs). TPPs serve as the gateway to payment services, for the end users.
    \item End users: these are individuals, businesses, government departments that are involved in transactions and pay money from or receive money to their bank accounts.
\end{enumerate}
TPPs will access Open Banking APIs via the MTS-OB API to offer CorDapp-enabled services utilising DLT-based smart contracts and payment flows. The MTS-OB layer handles the API calls and responses on behalf of the TPP, following end users’ consent (see Figure \ref{fig:openbankingtomtsobinteraction}). End users’ personal information data are kept securely in MTS-OB and each TPP can only access their customers’ transaction history.

\subsubsection{Core Banking System Layer}
The Core Banking System Layer of MTS-OB hosts the banks and financial institutions that hold deposit accounts for consumers, businesses, and government departments. This layer also holds the payment infrastructure which is regulated and maintained by the UK’s payment authority i.e. Payment Systems Regulator (PSR). Any payment issued via MTS-OB is ultimately settled via an interbank system which is part of the Core Banking System, such as Bankers' Automated Clearing System (Bacs), Clearing House Automated Payment System (CHAPS) and Faster Payment Service (FPS). Within MTS-OB, this layer remains unchanged, therefore no intervention is required by banks and payment systems regulators and operators to incorporate the new functionalities.

\subsubsection{Open Banking Layer}
The Open Banking API layer is where banks from the Core Banking System layer present bank account holder’s information to third parties. Each bank exposes a standard set of OBL compliant API endpoints which is the specific code instructions that an API user can invoke to cause a certain behaviour such as query a database. In MTS-OB, the Open Banking APIs are accessible from TPPs through the MTS DLT, where the payments are managed by Corda flows and smart contracts. The principal benefit of MTS-OB is there is no requirement for amending the Open Banking standard.

\subsubsection{MTS-OB and TPP Layer}
The MTS-OB layer is the heart of the MTS-OB architecture. Any payment service provider that participates in this layer, benefits from a portfolio of innovative services such as VSPs. Within the context of MTS-OB, the TPPs that provide MTS-enabled payment services run a Corda node in the MTS-OB DLT network. That node acts as the gateway to the Open Banking API’s layer and also contains the payment information that is used for VSPs and programmable money services. This information is shared with HMRC (via a dedicated HMRC Corda node) for fulfilling VSPs. Special purpose nodes called Notaries are also participating in the network, their sole purpose being to notarise transactions. With the MTS-OB in operation, the TPP layer is therefore integrated within MTS-OB (see Figure \ref{fig:mtsoblayers} for a full illustration of the layer participants). TPPs may be licensed by HMRC to provide VSP services to end users, or, alternative licensing schemes which are outside the scope of this paper.

\subsubsection{End User layer}
\begin{figure}
    \centering
    \includegraphics[width=1\linewidth]{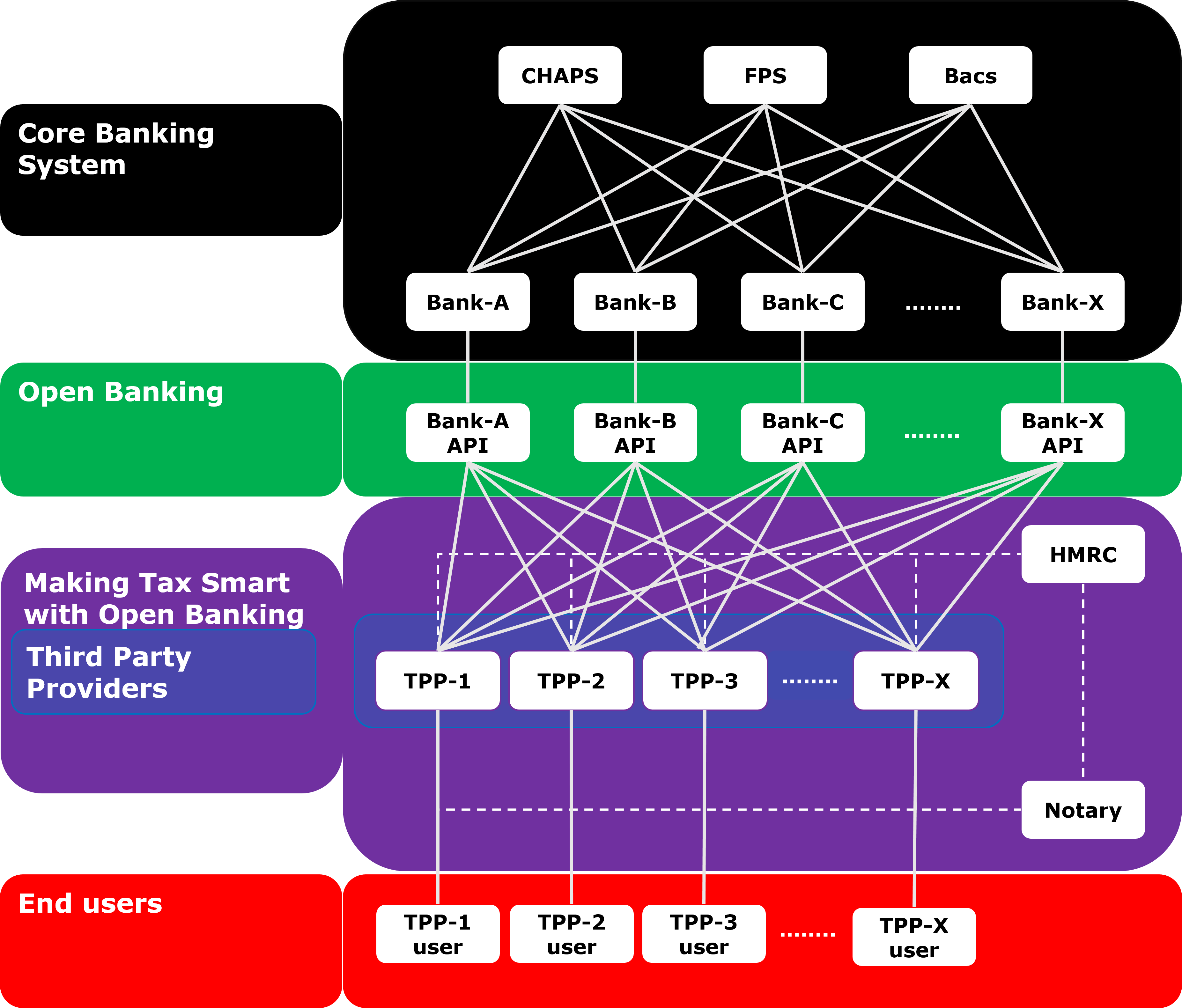}
    \caption{MTS-OB layers}
    \label{fig:mtsoblayers}
\end{figure}

This layer includes the end users of MTS. Consumers, small businesses that sell goods and provide services, large corporations that are involved in retail are members of this layer. The members benefit from the services offered by TPPs that operate within MTS-OB and access those services via web, or a mobile app. A summary of all MTS-OB layers is presented in Figure \ref{fig:mtsoblayers}.

\subsection{MTS-OB Layer on R3 Corda}
The R3 Corda is a permissioned and private distributed ledger technology platform, specifically designed for financial industry applications and with smart contract support \cite{valenta2017comparison}. In contrast to other DLT frameworks, transactions in Corda are only announced to relevant parties (need-to-know), with all communication in a Corda network taking the form of small multiparty protocols, called flows.

The unique permission and access-control management adopted in Corda is aligned with the MTS scheme where payments become trilateral instead of bilateral transactions. Each organisational entity that is involved in a payment participates by means of a Corda \textit{node}. This node contains a copy of the distributed ledger and the CorDapp; this is the distributed application that runs on the Corda platform and allows nodes to reach agreements concerned with updates on the ledger.

One major difference between Corda and other permissioned DLT platforms is the adoption of Notary nodes. These are special purpose dedicated hosts which participate in the network and serve the purpose of time-stamping and preventing double-spends of funds. Notaries are not owned or controlled by any member of the network. They are of special importance to MTS, as they serve as an independent authority which ensures fair participation from all entities and embodies trust in the system.
A new event, i.e. transferring money between accounts constitutes an on-ledger event that is called a Corda state. This is the fundamental data block stored on the Corda ledger and it constitutes an immutable record of a shared and agreed upon fact. When that fact has to be "updated", a new state is recorded, rendering the existing one as historic. The lifecycle of that fact over time is called a state sequence. In MTS-OB states represent paid and unpaid invoices. Corda contracts are rulesets that define how a Corda state can be updated to a new one. A contract is the "control" which ensures that a transaction has been fulfilled according to the pre-defined conditions. If the conditions are fulfilled, the state update can take place and record itself on the ledger. If the conditions are not fulfilled, the state update is not permitted and the corresponding transaction fails. In MTS-OB, contracts ensure that various conditions associated with the payment are met (e.g. that the VAT paid for an invoice matches the correct amount given the net value and VAT rate for the class of good sold), acting as a tax compliance actuator. Corda flows automate business processes by invoking a series of actions that cause state updates in a single execution. Actions such as networking, I/O and using sources of randomness are executed within a flow. More details about states, contracts and flows in MTS-OB are given in subsequent sections.

\subsubsection{MTS-OB on Corda: Nodes and Accounts}
TPPs and HMRC are the participants of the MTS-OB scheme, therefore they maintain a Corda node in the network. In the MTS-OB prototype, the Seller node is owned by a TPP and hosts multiple accounts for retail businesses. A Buyer node which is owned by another TPP hosts accounts for consumers. The HMRC node which hosts the facility for collecting the VAT from a split payment, hosts an account called VATpayments. Corda supports the use of accounts within a node, which act as logical partitions of the state objects which belong to that node. This enables hosting multiple entities within a single node and eliminating the need for each entity to host its own proprietary node. The Corda library accounts is used for two purposes: recording transactions for a particular entity (buyer, seller) and implementing permissioned data access control. End users of MTS-OB are assigned a Corda account, so that each end user’s transaction trail is isolated from the other users to prevent eavesdropping of personal data.

\begin{figure}
    \centering
    \includegraphics[width=1\linewidth]{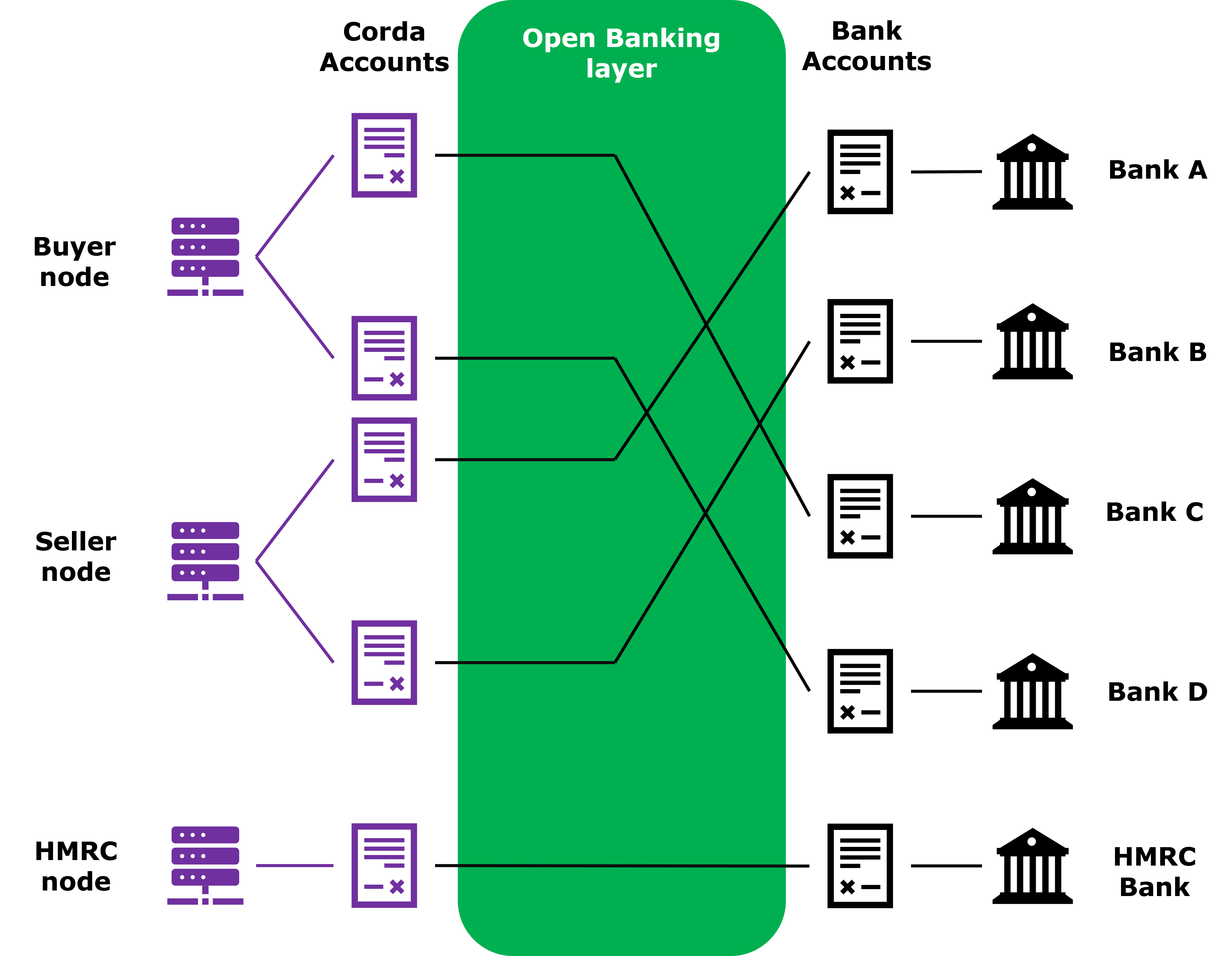}
    \caption{Linking accounts between Corda and Banks}
    \label{fig:mts2core}
\end{figure}

For each of the accounts within Corda, a matching account in a Bank outside the MTS-OB network (i.e. within the Core Banking layer in Figure \ref{fig:mtsoblayers}) must exist for the actual transfer of funds to take place. Each Corda account for a buyer or a seller is matched exactly to a bank account in any one of the Banks in the Core Banking layer (see Figure \ref{fig:mts2core}) The mapping between Corda accounts and Bank accounts is held in a separate database.

\subsubsection{MTS-OB on Corda: States}
The key Corda state employed in MTS-OB is called SMInvoiceState and is defined in a Kotlin file. This state holds all the information concerned with an invoice and the associated transfer of funds. The class SMInvoiceStateOBP specifies the properties of an invoice state (see Table \ref{tab:sminvoicestate}) and the function calculateVAT()for calculating the VAT payable given a shopping list which is a property of the state. The property goodsPurchased is of the type ShoppingListItems which is defined as shown in the following code snippet:

\begin{lstlisting}[caption={goodsPurchased data type}, label={datatype}]
data class ShoppingListItems (
    val item: String,
    val quantity : Int,
    val price : Long,
    val vatRate : Double
)
\end{lstlisting} 
\begin{table}[h]
\caption{List of Properties for Class \texttt{SMInvoiceStateOBP}}
\label{tab:sminvoicestate}
\begin{tabular}{p{0.25\linewidth} p{0.65\linewidth}}
\hline
\textbf{Property Name} & \textbf{Description} \\
\hline
\texttt{amount} & The total amount for the invoice. \\
\texttt{issuedDate} & The date that the invoice was issued. \\
\texttt{amountPaid} & The amount that has been paid (equal to 0 if invoice is unpaid). \\
\texttt{paidFrom} & The Party that issued the currency for paying this invoice. \\
\texttt{Seller} & The seller account specified as \texttt{AnonymousParty} (only the information required for identifying a Party while excluding private information). \\
\texttt{Buyer} & The buyer account specified as \texttt{AnonymousParty} (only the information required for identifying a Party while excluding private information). \\
\texttt{Hmrc} & The HMRC account specified as \texttt{AnonymousParty} (only the information required for identifying a Party while excluding private information). \\
\texttt{goodsPurchased} & The list of items bought specified as \texttt{ShoppingListItems} object. \\
\texttt{vatPaid} & The VAT that has been paid. \\
\texttt{paidDate} & The date that the invoice was paid. \\
\texttt{netAmountReceived} & The amount that the seller has received after VAT has been paid. \\
\texttt{OBPTransactions} & A list of the transaction IDs for the matching payments through Open Banking. \\
\hline
\end{tabular}
\end{table}

\subsubsection{MTS-OB on Corda: Contracts}
The main Corda contract employed in MTS-OB is the \texttt{InvoiceContractOBP} defined in the file \texttt{InvoiceContractOBP.kt}. This contract specifies the requirements for an \texttt{InvoiceState} to be valid and accepted by all parties that are involved in the flow. If any of the requirements is not satisfied, then the transaction fails in the Corda node that it is being executed in and the corresponding State is not created (if the transaction is to create a new invoice) or updated (if the transaction is to pay for an invoice). The contract is summarized in the pseudocode presented in Algorithm \ref{contract}.

\begin{algorithm}
\caption{Contract Verification Function}
\label{contract}
\begin{algorithmic}[1]

\Function{verify}{$transaction$}
    \State $command \gets transaction.\textsc{getCommand}()$
    \Switch{$command$}
        \Case{\textit{Issue}}
            \Comment{Constraints for issuing an invoice}
            \State \textbf{ensure} $transaction.inputs.\textsc{isEmpty}()$
            \Comment{No inputs consumed when issuing an Invoice}
            \State \textbf{ensure} $transaction.outputs.size = 1$
            \Comment{Only one output state should be created}
            \State $outputInvoice \gets transaction.\textsc{getOutput}(InvoiceState)$
            \State \textbf{ensure} $outputInvoice.buyer \neq outputInvoice.seller$
            \Comment{The seller and the buyer cannot be the same}
            \State \textbf{ensure} $outputInvoice.amount > 0$
            \Comment{The Invoice's value must be a positive number}
            \State \textbf{ensure} $outputInvoice.amountPaid = 0$
            \Comment{The Invoice should not be a paid invoice}
        \EndCase

        \Case{\textit{Pay}}
            \Comment{Constraints for paying an invoice}
            \State $restrictedGoods \gets \textsc{load}(\texttt{"/restrictedGoods.txt"})$
            \Comment{Load restricted goods list}
            \State \textbf{ensure} $transaction.\textsc{input}(InvoiceState).size = 1$
            \Comment{One invoice state is consumed}
            \State $groupedInvoices \gets transaction.\textsc{groupStates}(InvoiceState)$
            \State \textbf{ensure} $groupedInvoices.inputs.size = 1$
            \Comment{There must be one input Invoice}
            \State $inputInvoice \gets transaction.\textsc{getInput}(InvoiceState)$
            \State \textbf{ensure} $inputInvoice.amountPaid = 0$
            \Comment{The Invoice must not have been paid}
            \State $outputInvoice \gets transaction.\textsc{getOutput}(InvoiceState)$
            \State \textbf{ensure} $outputInvoice.amount = outputInvoice.amountPaid$
            \Comment{Amount paid must equal the amount in Invoice}
            \State \textbf{ensure} $inputInvoice.buyer = outputInvoice.buyer$
            \Comment{Buyer pays an invoice issued to them}

            \If{$outputInvoice.amountPaid.issuer = \texttt{"HMRC"}$}
                \For{$item \textbf{ in } outputInvoice.goodsPurchased$}
                    \State \textbf{ensure} $item.name \notin restrictedGoods$
                    \Comment{You may not purchase $item$ from this account}
                \EndFor
            \EndIf
        \EndCase
    \EndSwitch

    \State $signers \gets transaction.\textsc{getCommandSigners}()$
    \State $participantKeys \gets transaction.\textsc{getOutputStateParticipants}()$
    \State \textbf{ensure} $signers.\textsc{containsAll}(participantKeys) \wedge signers.size = 3$
    \Comment{All (supplier, buyer, and HMRC) must sign Invoice Payment transaction}
\EndFunction

\end{algorithmic}
\end{algorithm}

\subsubsection{MTS-OB on Corda: Flows}
The processes of issuing and paying for an invoice in MTS-OB are implemented with the use of Corda flows and are specified in the files \texttt{IssueInvoiceOBP.kt} and \texttt{PayInvoiceOBP.kt}. These flows initialise or modify, as necessary, the cryptographic keys of the counterparties and the field values that are later placed within the invoice state. They also ensure that all interested parties have signed the transaction. In the case of the \texttt{PayInvoiceOBP} flow, there is added functionality which calls the appropriate Open Banking endpoints to i) check that the payer has sufficient funds for the transaction i.e. the Account Information Service \cite{obp2024ais} and ii) to initiate the payments for the net amount payable for the goods sold and for the VAT payable to HMRC i.e. the Payment Initiation Service \cite{obp2024pis}.

\paragraph{IssueInvoice Flow}

\begin{figure}
    \centering
    \includegraphics[width=1\linewidth]{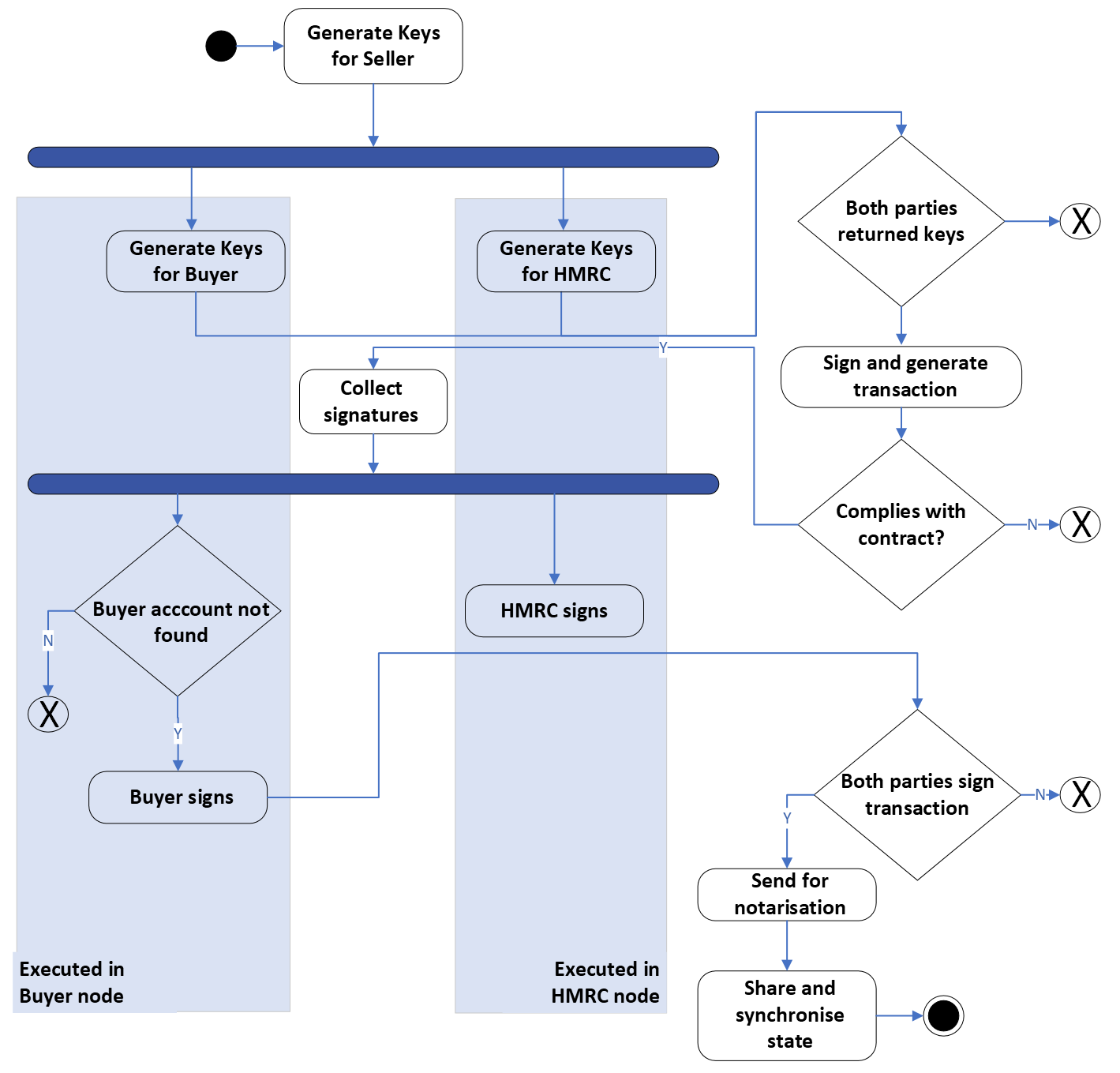}
    \caption{Activity Diagram for the IssueInvoice flow}
    \label{fig:issueactivitydiagram}
\end{figure}

\begin{figure}
    \centering
    \includegraphics[width=1\linewidth]{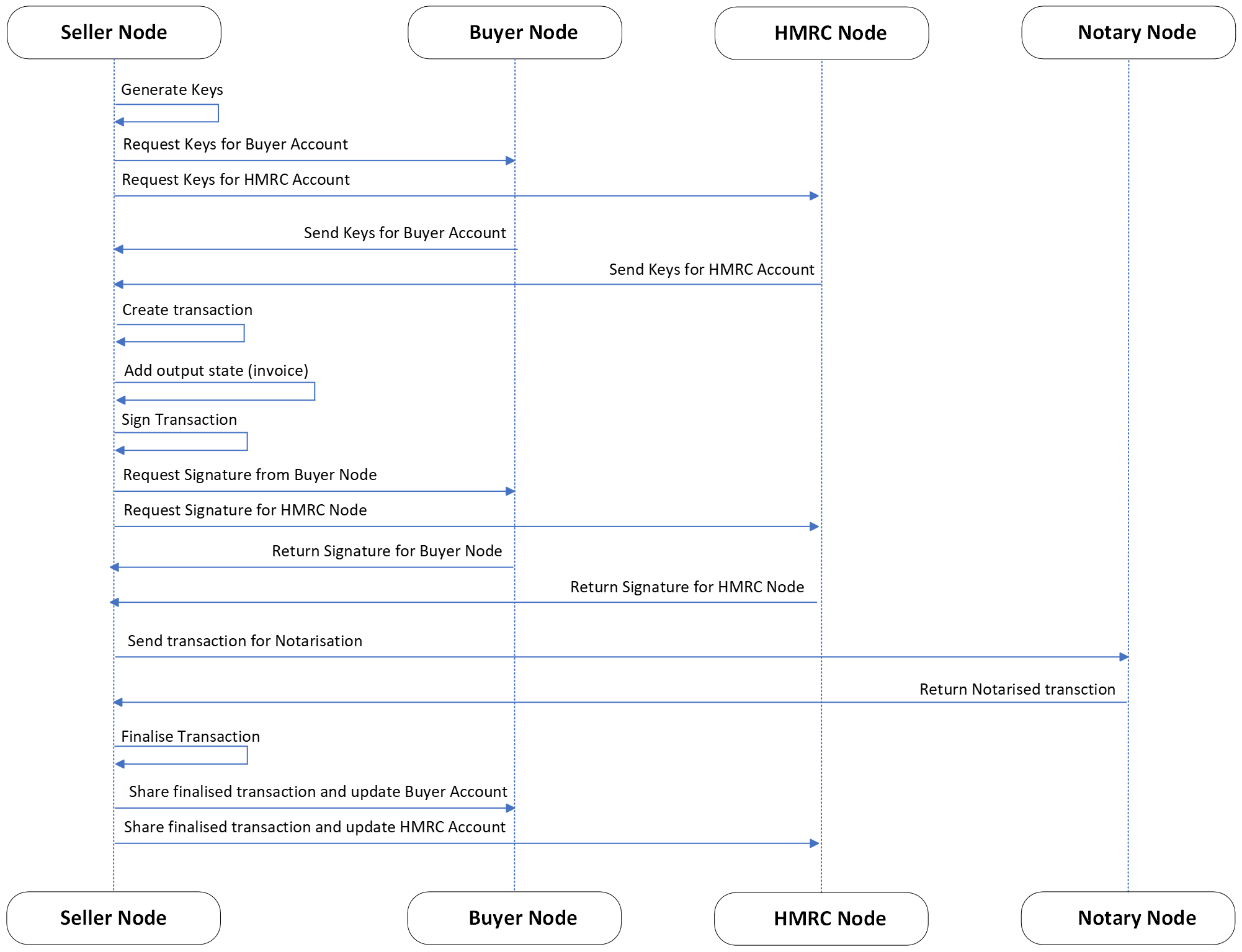}
    \caption{Sequence Diagram for the IssueInvoice flow}
    \label{fig:issuesequencediagram}
\end{figure}

\begin{figure}
    \centering
    \includegraphics[width=1\linewidth]{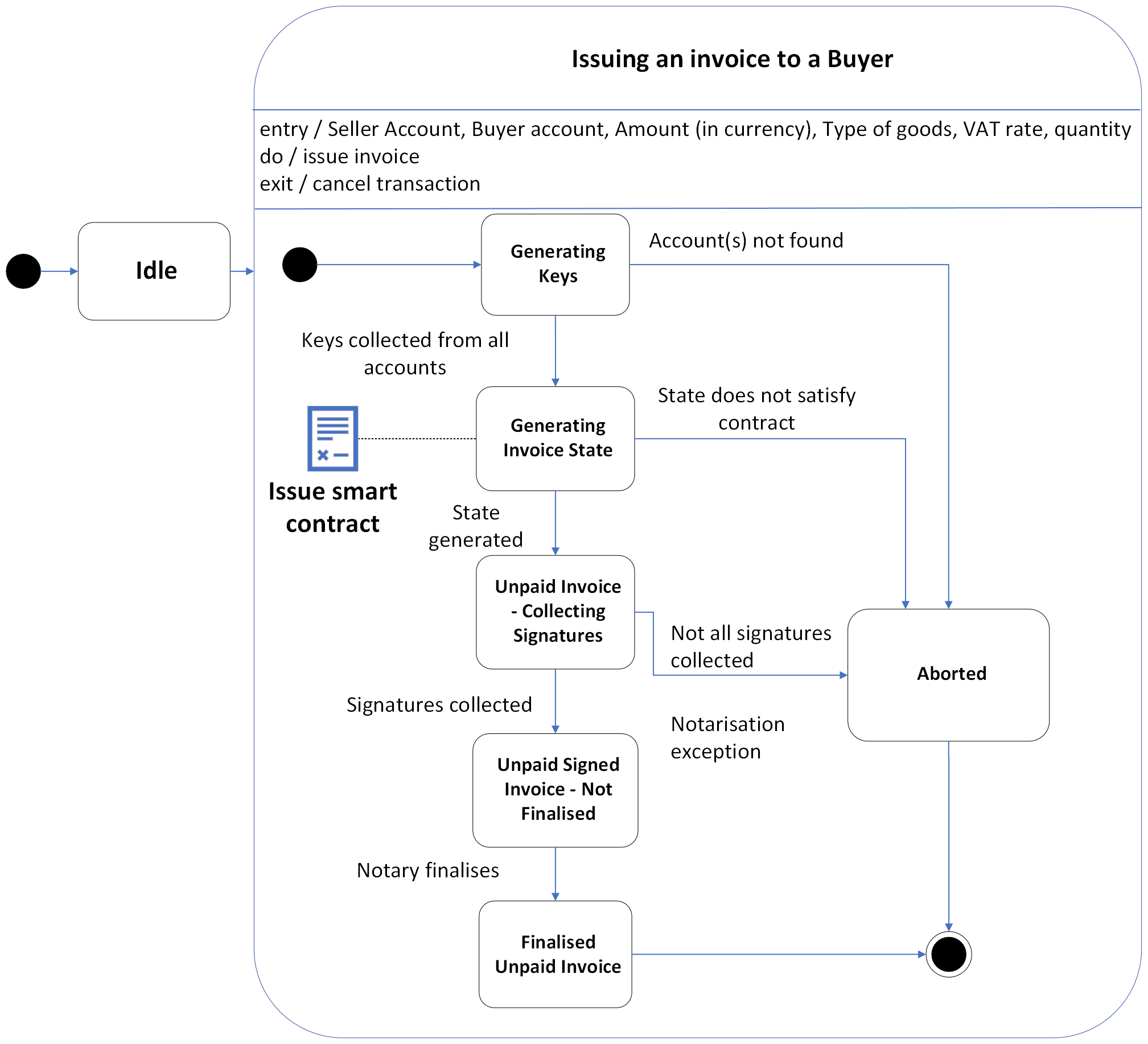}
    \caption{State Diagram for the IssueInvoice flow}
    \label{fig:issuestatediagram}
\end{figure}

The \texttt{IssueInvoiceOBP} flow involves the following steps on behalf of the \texttt{Seller}: i) generating one-time keys for each account involved in the process, ii) generating the transaction, iii) collecting signatures for the transaction from all the accounts involved and iv) creating the invoice state, checking it against the smart contract and sharing the state with the counterparties. The counterparties, i.e. \texttt{Buyer} and \texttt{HMRC} check the transaction shared with them by the Seller against the \texttt{InvoiceContractOBP} and if the conditions in the contract are fulfilled, they proceed with signing the transaction and notify the \texttt{Seller}. Then, the \texttt{Seller} shares the finalised \texttt{InvoiceState} with \texttt{Buyer} and \texttt{HMRC} and submits the fully signed transaction to the \texttt{Notary} node. The \texttt{Notary} validates the timestamp of the transaction, notifies the interested parties to update their ledgers with the new \texttt{InvoiceState} and returns a unique transaction ID for the completed payment. When the transaction is completed, the ledger contains a new state which specifies an unpaid invoice which is only visible to the interested parties. The flow is illustrated by UML Activity, Sequence and State diagrams in Figures \ref{fig:issueactivitydiagram}, \ref{fig:issuesequencediagram}, and\ref{fig:issuestatediagram}. 

\paragraph{PayInvoice Flow}
For the payment side, the purchaser is informed of the pending payment and invoice through their payment interface (e.g. a smartphone or web banking app). After reviewing the details of the payment and agreeing to the terms, the \texttt{PayInvoiceOBP} flow is executed. First, the \texttt{Buyer} performs the same steps performed by the Seller in the \texttt{IssueInvoiceOBP} flow i.e. i) generates one-time keys for each account involved in the process, ii) generates the transaction, iii) collects signatures for the transaction from all the accounts involved iv) creates the invoice state, checks it against the smart contract and shares the state with the counterparties. For the payment side, the \texttt{Buyer} queries the mappings database for discovering which Bank accounts that are accessible via the Open Banking API correspond to the \texttt{Buyer}, \texttt{Seller} and \texttt{HMRC} accounts in Corda. This is followed by calculating the amounts to be paid, i.e. the net amount to the \texttt{Seller} account and the VAT amount to \texttt{VATPayments} account in HMRC. The flow then issues an AIS query via the appropriate Open Banking API endpoint and transmits it to the \texttt{Buyer} bank Open Banking API which returns the account balance. If that balance is sufficient for the full amount of the payment, a new \texttt{InvoiceState} is created and a PIS request is sent to the buyer’s Bank Open Banking API, using the appropriate Open Banking API endpoints for transferring money. The node then builds a transaction for payment, signs it and initiates sessions with the interested parties who are involved in the transaction.

\begin{figure}
    \centering
    \includegraphics[width=1\linewidth]{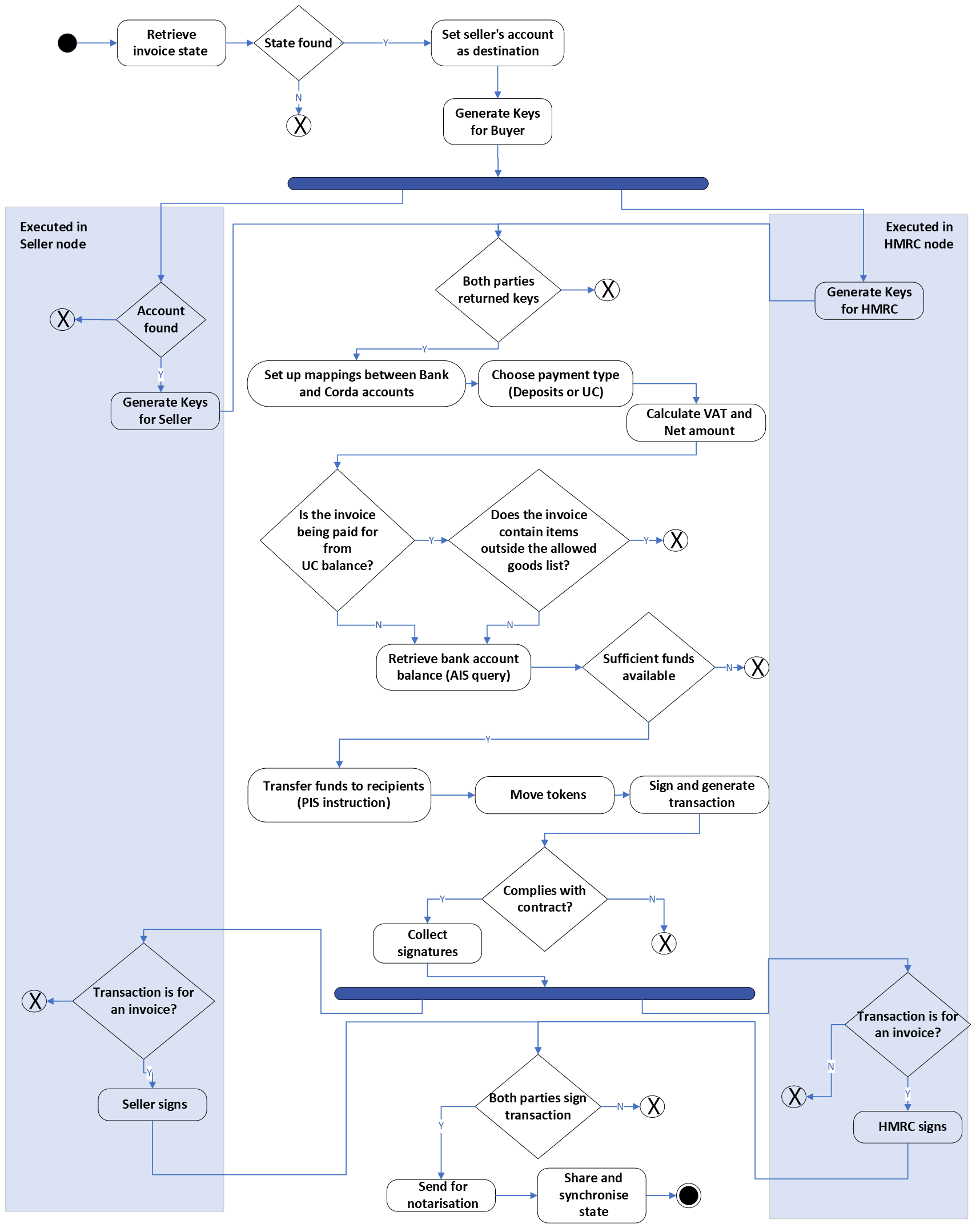}
    \caption{Activity Diagram for the PayInvoice flow}
    \label{fig:payactivitydiagram}
\end{figure}

\begin{figure}
    \centering
    \includegraphics[width=1\linewidth]{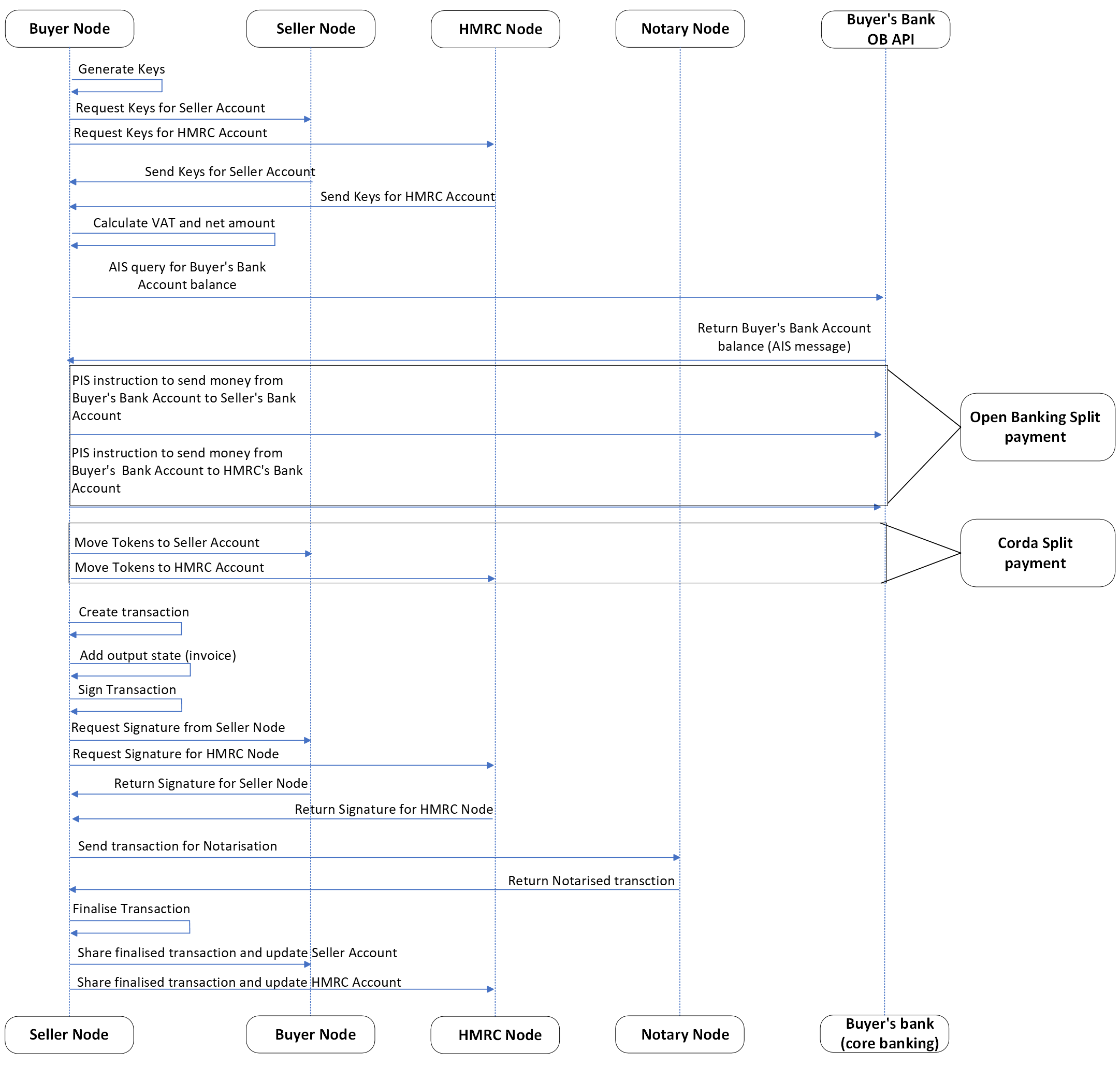}
    \caption{Sequence Diagram for the PayInvoice flow}
    \label{fig:paysequencediagram}
\end{figure}

\begin{figure}
    \centering
    \includegraphics[width=1\linewidth]{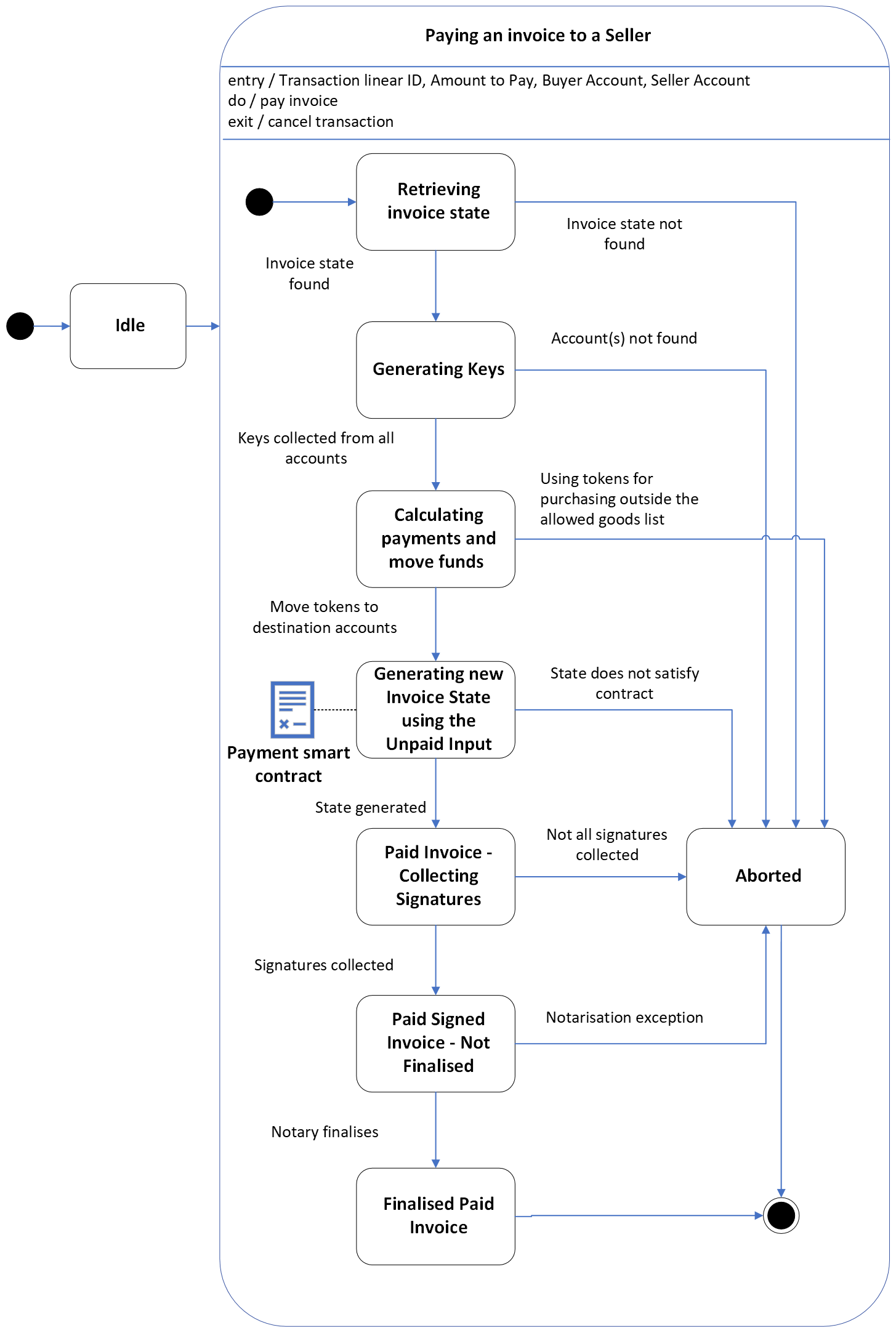}
    \caption{State Diagram for the PayInvoice flow}
    \label{fig:paystatediagram}
\end{figure}

The interested parties, i.e. \texttt{Seller} and \texttt{HMRC} then check the transaction against the \texttt{InvoiceContractOBP} and if the conditions in the contract are fulfilled, they proceed with signing the transaction and notify \texttt{Buyer}. Then, \texttt{Buyer} shares the finalised \texttt{InvoiceState} with \texttt{HMRC} and \texttt{Seller} and submits the fully signed transaction to the \texttt{Notary} node. The \texttt{Notary} validates the timestamp of the transaction, notifies the interested parties to update their ledgers with the new \texttt{InvoiceState} state and returns a unique transaction ID for the completed payment. When the transaction is completed, the ledger contains a new state which specifies a paid invoice which is visible to the interested parties. The old \texttt{InvoiceState} state for the unpaid invoice is now unusable and that invoice cannot be paid again. The flow is illustrated by UML Activity, Sequence and State diagrams in Figures \ref{fig:payactivitydiagram}, \ref{fig:paysequencediagram}, \ref{fig:paystatediagram}. 

\subsubsection{Open Banking endpoints}
The Open Banking API endpoints that are used by participants in the UK fintech ecosystem are dictated by the OBL and are defined in the Open Banking API Specifications \cite{obie2024apispecs}. Three API endpoints are required for MTS-OB to function:
\begin{enumerate}
    \item \url{/my/logins/direct}: Used from the TPP to getting a token from Open Banking and use it to make authenticated API calls in the future i.e. the calls specified below. 
    \item \url{/banks/bankID/accounts/accountId/balances}: Used from the TPP to query bank \texttt{bankID} about the balances of account \texttt{accountID}. This enables the Corda flow to discover if a payer has sufficient funds available for fulfilling a payment request. 
    \item \url{/banks/bankID/accounts/fromAccountId/owner/transaction-request-types/ACCOUNT/transaction-requests}: Used from the TPP to request a transfer of funds from the account \texttt{fromAccountID}. This enables the Corda flow to discover if a payer has sufficient funds available for fulfilling a payment request. The remaining details, such as payee’s bank and account, currency, value and description are contained in the transaction body in the following format:

\begin{lstlisting}[caption={Pay Invoice transaction body format}, label={txbodyformat}]
"to":{"bank_id":"<bank_id>","account_id":"<account_id>"},
"value":{"currency":"<currency>","amount":"<amount>"},
"description":"<description>"
\end{lstlisting} 
Where: 
\begin{itemize}
    \item \texttt{bankID} is the ID of the payee’s bank 
    \item \texttt{accountID} is the ID of the payee’s account
    \item \texttt{currency}: the currency for the payment
    \item \texttt{amount}: the amount to be paid
    \item \texttt{description}: a String field for providing a free text description of the payment
\end{itemize}
\end{enumerate}

\section{Experimental Design and MTS-OB prototype}
\subsection{Performance factors for measuring scalability of DLT applications}
DLT-based applications for a multitude of domains are currently being researched, with a focus on demonstrating the feasibility and scalability of these solutions. DLTs which are based on either permissioned or permissionless access models are assessed with respect to how certain application requirements affect several performance metrics of their applications. Feasibility studies on the application of DLTs in particular, are concerned with how certain operating factors affect a number of performance metrics. 
As part of the research presented herein, a literature review has been conducted which shows that the majority of studies emphasise on throughput and latency as there is a common trade-off between data integrity and processing speed for DLT solutions. Of particular interest is the capability of a DLT application to scale with increasing volumes of workload. Workload is assessed in terms of how many transactions are submitted to the ledger per unit time \cite{jiang2020performance}, \cite{huang2020scalable}, \cite{melo2019using}, \cite{khan2020investigating}, \cite{novo2018blockchain}, \cite{hao2018performance}, 
\cite{han2025layer2}, 
\cite{liang2026distributed},\cite{kuzlu2019performance}, \cite{pal2020design}, \cite{sharma2018blockchain}, as well as the size of the data which is processed for each transaction \cite{khan2020investigating}, \cite{novo2018blockchain}, \cite{han2025layer2}, \cite{liang2026distributed}, \cite{sharma2018blockchain}, \cite{wang2019blockchain}, \cite{si2019iot}, \cite{lu2019ubaas}, \cite{chen2019design}, \cite{li2019crowdbc}. Other works study the effect of the employed consensus mechanism (Proof-of-Work, Proof-of-Stake, Byzantine Fault Tolerance) \cite{khan2020investigating}, \cite{novo2018blockchain}, \cite{hao2018performance}, \cite{kuzlu2019performance}, \cite{sharma2018secure}. The number of participating entities can also have an effect in performance \cite{novo2018blockchain}, \cite{kuzlu2019performance}, \cite{sharma2018blockchain} , \cite{wang2019blockchain}, \cite{si2019iot} , \cite{lu2019ubaas}, \cite{li2019crowdbc}, as more nodes and users present a delimiting effect in transaction throughput \cite{jiang2020performance}, \cite{melo2019using}, \cite{pal2020design}, \cite{si2019iot}, \cite{lu2019ubaas} , \cite{zhou2020solutions} and latency \cite{jiang2020performance}, \cite{huang2020scalable}, \cite{melo2019using}, \cite{khan2020investigating}, \cite{novo2018blockchain}, \cite{hao2018performance}, \cite{pal2020design}, \cite{sharma2018blockchain}, \cite{si2019iot}, \cite{lu2019ubaas}, \cite{wang2019blockchainprivacy} It has also been found that when the number of transactions becomes very high, some of those transactions might fail, it is therefore useful to measure the loss rate as a performance metric \cite{jiang2020performance}, \cite{huang2020scalable}, \cite{melo2019using}, \cite{khan2020investigating}, \cite{zhou2020solutions}. With reference to the review outlined above, the performance factors metrics and metrics employed for assessing the scalability of DLT solutions are summarised in Table \ref{tab:factors}. 

\begin{table}[h]
\caption{Factors and Metrics in the Literature}
\label{tab:factors}
\begin{tabular}{p{0.45\linewidth} p{0.45\linewidth}}
\hline
\textbf{Factor} & \textbf{Source} \\
\hline
Transaction Load & {\cite{jiang2020performance}, \cite{huang2020scalable}, \cite{melo2019using}, \cite{khan2020investigating}, \cite{novo2018blockchain}, \cite{hao2018performance}, \cite{kuzlu2019performance}, \cite{pal2020design}, \cite{sharma2018blockchain}} \\
Data Volume & {\cite{khan2020investigating}, \cite{novo2018blockchain}, \cite{sharma2018blockchain}, \cite{wang2019blockchain}, \cite{si2019iot}, \cite{lu2019ubaas}, \cite{chen2019design}, \cite{li2019crowdbc}} \\
Consensus Mechanism (*) & {\cite{khan2020investigating}, \cite{novo2018blockchain}, \cite{hao2018performance}, \cite{kuzlu2019performance}, \cite{sharma2018secure}} \\
No. of Clients & {\cite{khan2020investigating}, \cite{novo2018blockchain}, \cite{kuzlu2019performance}, \cite{sharma2018blockchain}, \cite{wang2019blockchain}, \cite{si2019iot}, \cite{lu2019ubaas}, \cite{li2019crowdbc}} \\
\hline
\textbf{Metric} & \textbf{Source} \\
\hline
Throughput & {\cite{jiang2020performance}, \cite{melo2019using}, \cite{pal2020design}, \cite{si2019iot}, \cite{lu2019ubaas}, \cite{zhou2020solutions}} \\
Latency & {\cite{jiang2020performance}, \cite{huang2020scalable}, \cite{melo2019using}, \cite{khan2020investigating}, \cite{novo2018blockchain}, \cite{hao2018performance}, \cite{pal2020design}, \cite{sharma2018blockchain}, \cite{si2019iot}, \cite{lu2019ubaas}, \cite{wang2019blockchainprivacy}} \\
Transaction Rejection Rate (*) & {\cite{jiang2020performance}, \cite{huang2020scalable}, \cite{melo2019using}, \cite{khan2020investigating}, \cite{zhou2020solutions}} \\
\hline
\end{tabular}
\end{table}

The study presented in this paper investigates the impact of two factors on performance: Transaction Load and Number of Clients. Transaction Load $tx$ is the number of invoices submitted for issue or payment by each client in a run. Number of Clients $cl$ is the number of endpoints that communicate with a participating node and submit transactions. Data Volume, i.e. the number of line items carried on an invoice, is not varied in the present study. In the antecedent Smart Money CBDC work \cite{louvieris2024making}, which uses the same CorDapp architecture, this factor was examined at three levels — 10, 50 and 100 items per invoice, at a fixed total of 1{,}000 invoices — and produced no separation within the families of curves for either issuance or payment, establishing that the number of items per invoice has no discernible effect on throughput or latency. That finding is adopted here and the item count held constant at 10 across all runs.
 
Corda does not employ a consensus algorithm in the sense used by \cite{hao2018performance} and \cite{sharma2018secure}; it provides a reliable witness by means of Notary nodes which 
prevent double-spending \cite{r3corda2023}. The consensus mechanism is therefore excluded from the design. Transaction Rejection Rate is not reported as a metric because it was zero at every factor level: across all 2{,}000 runs, every submitted transaction completed successfully. The factors and metrics carried forward are summarised
in Table~\ref{tab:factorsdesc}.

Within the context of a CBDC, the term "endpoints" may refer to: a) end-user devices which hold an account holder's private key (plastic card, mobile phone) and interact directly with the TPP’s Corda node or b) a special-purpose "aggregator" server which is operated by a TPP and collects transaction requests from end-users and submits them to the Corda node in bulk. By varying each of the factors (which constitute the independent variables of the experiment), the effect on scalability can be measured and analysed. Regarding the effect of the consensus mechanism on scalability, Corda does not use a consensus system. Instead, it provides a reliable witness, by means of Notary nodes which prevent double-spending \cite{r3corda2023}. Therefore, the impact of the consensus mechanism is not considered in the experiment.
With reference to the performance metrics, results for the variable Transaction Rejection Rate are not presented, as Corda has consistently completed all the transactions throughout the tests conducted. Therefore, only Throughput and Latency have been selected as the measurement variables for the tests. The performance factors and metrics are summarised in Table \ref{tab:factorsdesc}.

\begin{table}[h]
\caption{Factors and Metrics}
\label{tab:factorsdesc}
\begin{tabular}{p{0.30\linewidth} p{0.60\linewidth}}
\hline
\textbf{Factor} & \textbf{Description} \\
\hline
Transaction Load $tx$ & The number of transactions submitted to the ledger. \\
No. of Clients $cl$ & The number of clients that interact with the ledger via a correspondent node. \\
\hline
\textbf{Metric} & \textbf{Description} \\
\hline
Throughput $tr$ & The number of successful transactions completed per unit time. Measured in Transactions per Second (TPS). \\
Latency $la$ & The time delay between two events: 1) transaction issued and 2) transaction validated. Measured in seconds. \\
\hline
\end{tabular}
\end{table}

\subsection{MTS-OB Prototype}
To test the scalability of the MTS-OB architecture, an OB prototype has been developed and tested in a controlled experimental environment. The aim of the tests is to demonstrate that the MTS with Open Banking approach scales with increasing workloads therefore can support VSPs in a real-world application. The prototype consists of the following components:
\begin{enumerate}
    \item The Open Banking Sandbox.
    \item The MTS-OB architecture implemented on R3 Corda.
    \item The Mappings database between Corda and Bank accounts.
    \item The RPC clients for simulating end users (i.e. buyers and sellers) of the VSP service.
\end{enumerate}

\subsubsection{Open Banking Sandbox}
The Open Banking Sandbox simulates the functionality of the OBL directory and a reference back-end Core Banking system. The prototype adopts OBL standards that are in accordance with the Payment System Regulator (PSR) and Financial Conduct Authority (FCA) regulations in place, therefore is a compliant infrastructure capability for integrating with the UK’s payment system. It provides the required functionality for a proof-of-concept which proves that MTS-OB is a viable application for enabling VSPs without posing additional risks and disrupting the money system.
For the purposes of the research presented herein, the open source OpenBankProject/OBP-API RESTful API platform v.4.0.0 was selected as for running a sandbox within a controlled test environment. This option was chosen over corporate Open Banking sandboxes (e.g. HSBC, NatWest, Barclays) because they provide inadequate facilities, i.e. a limited number of bank branches and accounts per branch, to assess the scalability of MTS-OB. On the other hand, the OpenBankProject/OBP-API provides the required API endpoints for creating unlimited number of bank branches and accounts, and for allocating umlimited funds to each account.

The simulated core banking system is represented as an H2 database, which consists of two files: (\texttt{lift\_proto.db.mv.db} and \texttt{lift\_proto.db.trace.db}) and stores the following information about the core banking system:

\begin{enumerate}
    \item Banks
    \item Bank accounts per bank (with balances)
    \item Transaction history between banks
    \item API calls history
\end{enumerate}
The database is accessible via Java Web App which is based on the Scala plugin.

\subsubsection{MTS-OB on Corda}
MTS has been implemented in R3 Corda version 4.8 (see section MTS-OB layer on R3 Corda for the architecture) which was the latest version at the time of development. There are four Corda nodes in total for the MTS-OB prototype (see Table \ref{tab:cordanodes}). Each node resides within a Java Virtual Machine (JVM) and has a copy of the contracts and flows written for MTS-OB. The RPCClients are executed from the machine that hosts the JVMs.

\begin{table}[h]
\caption{Corda Nodes in the MTS-OB Prototype}
\label{tab:cordanodes}
\begin{tabular}{p{0.15\linewidth} p{0.25\linewidth} p{0.50\linewidth}}
\hline
\textbf{Node Name} & \textbf{Accounts in Node} & \textbf{Description} \\
\hline
\texttt{HMRC} & \texttt{VATPayments} & Operated by HMRC. It ensures that the correct VAT has been paid for a transaction. \\[6pt]
\texttt{Buyer} & 100 accounts for end users & Operated by a TPP which hosts accounts of consumers that make purchases. \\[6pt]
\texttt{Seller} & \texttt{MegaCompany}, \texttt{MediumCompany}, \texttt{MiniCompany}, \texttt{SmallBusiness}, \texttt{SoleTrader}, \texttt{ElectricAndGas} & Operated by a TPP which hosts accounts of businesses that make sales. \\[6pt]
\texttt{Notary} & No accounts & Essential Corda node which validates transactions before they are added to the ledger. \\
\hline
\end{tabular}
\end{table}

\subsubsection{Mappings database and RPC Client}
The Mappings database is a json file that holds the mappings between accounts in Corda and the OBP Sandbox. A code snippet from the file is given in Listing \ref{lst:mappingsdb}. Each json object contains: the name of the account in Corda, the type of account (Current or Universal Credit), the account ID for the OBP sandbox, the corresponding bank in the OBP sandbox and an IBAN. 

\begin{lstlisting}[language=json, caption={Account configuration file}, label={lst:mappingsdb}]
[ {
  "CordaAccountName" : "VATPayments",
  "Account" : [ {
    "type" : "Current",
    "obpaccountId" : "5e7e3d0a-6f53-4a47-9c05-8e88b2f7d69c",
    "obpbank" : "HMRCBank",
    "iban" : "GB1234567890010"
  } ]
}, {
  "CordaAccountName" : "Alfreda",
  "Account" : [ {
    "type" : "Current",
    "obpaccountId" : "24a5e67f-5d74-448e-b312-b7d9072a9135",
    "obpbank" : "Bank-C-UK",
    "iban" : "GB123456789001"
  }, {
    "type" : "UC Account",
    "obpaccountId" : "0a9425b0-3213-4e6c-8b3d-1130f7525870",
    "obpbank" : "Bank-C-UK",
    "iban" : "GB98765432001"
  } ]
}, {
  "CordaAccountName" : "Allissa",
  "Account" : [ {
    "type" : "Current",
    "obpaccountId" : "39f965e5-65b0-4937-9b01-e4a1c1b9456d",
    "obpbank" : "Bank-B-UK",
    "iban" : "GB123456789002"
  }, {
    "type" : "UC Account",
    "obpaccountId" : "ffe24674-1745-418f-862e-83b870008ca7",
    "obpbank" : "Bank-B-UK",
    "iban" : "GB98765432002"
  } ]
}, {
  "CordaAccountName" : "Allx",
  "Account" : [ {
    "type" : "Current",
    "obpaccountId" : "b619e5cb-3f03-4923-aa8d-6bf6582575ff",
    "obpbank" : "Bank-A-UK",
    "iban" : "GB123456789003"
  }, {
    "type" : "UC Account",
    "obpaccountId" : "cb98acbb-cf8f-4901-9c13-422293101da2",
    "obpbank" : "Bank-A-UK",
    "iban" : "GB98765432003"
  } ]]

\end{lstlisting}

An RPC (Remote Procedure Call) client application is used for interfacing an account with a node. That RPC is submitted to the target node through the network hence the user can reside in any network station that can communicate with the node through TCP/IP. Corda 4.8 is shipped a client app in thae form of a java archive named TestRPCClientShopping-0.1-all.jar.

\subsection{Experimental Design}
\label{sec:expdesign}
The purpose of the experiment is to demonstrate that MTS-OB is a scalable design for automated VSPs in the UK. We measure the effects caused to system performance by applying variable levels of stress (through the various factor levels) in order to demonstrate the design’s scalability. 

A full factorial design was used, in which every combination of the two factors was executed. Transaction Load $tx$ was varied over $\{10, 20, \ldots, 100\}$ invoices per client and Number of Clients $cl$ over $\{10, 20, \ldots, 100\}$ clients, giving 100 factor combinations with every cell populated. Each combination was executed once for invoice issuance and once for invoice payment, and each execution repeated 10 times, giving $100 \times 2 \times 10 = 2{,}000$ runs.
 
In the DLT performance literature a design is considered to scale when the cost of servicing additional workload grows no faster than the workload itself, so that the resources required to meet a target load can be projected from measurement \cite{lu2019ubaas, kuzlu2019performance, zhou2020solutions}. This requires that transaction latency grow at most linearly with offered load, that throughput degrade predictably rather than collapsing, and that no superlinear cost be introduced by increasing the number of participating endpoints. Superlinear latency growth, or throughput collapse under sustained load, would indicate an architectural bottleneck such as lock contention or quadratic message complexity, which would not be handled by allocating additional resources.

 
The measurements characterise a single MTS-OB node servicing a growing client population and do not measure capacity added by provisioning further nodes. MTS-OB is designed so that each TPP runs its own Corda node (or cluster of nodes) with its own client population. Adding TPPs therefore adds capacity, so the system's total capacity is not bounded by the single-node figures reported here. Second, all components of the testbed share one physical machine, so absolute figures reflect resource contention between the RPC clients, the four Corda nodes and the OBP sandbox. The trends, rather than the absolute values, are the object of the study.

\subsubsection{Testbed}
 
All components are executed on a single host: an Intel i5-12600K at 3.7\,GHz with 80\,GB RAM and a 1\,TB NVMe drive, running Windows 11, with Java 1.8 for Corda and Java 1.16 for the OBP API. Because every component is co-located, inter-component communication is over the loopback network interface and carries no physical network latency; conversely, the components contend for the same CPU, memory and I/O resources throughout.

The MTS-OB network comprises the four Corda nodes listed in
Table~\ref{tab:cordanodes}, each in its own JVM. The notary is a validating notary running as a single node. The prototype uses Corda Open Source 4.8, which runs a single flow worker per node; the implications of this are addressed in Section~\ref{sec:validity}. The Open Banking layer is an OpenBankProject/OBP-API v4.0.0 sandbox persisting to an in-process H2 database.
 
Factor ranges were bounded by the testbed. Each client is an independent
Windows PowerShell process running the Corda RPC client, occupying approximately 384\,MB of resident memory, so $cl = 100$ is the largest client population the host sustains alongside the Corda nodes and the OBP sandbox. Maximum total offered work is therefore 10{,}000 invoices issued and 10{,}000 paid per repetition.

\subsubsection{Instrumentation and procedure}
 
Client processes are launched by a batch script which spawns the required number of PowerShell windows in sequence. Once started, all clients execute concurrently for the remainder of the run. Latency is the elapsed time of the client-side code block that executes the issue or pay operation, timed from the start of that block to its completion, and therefore spans the complete flow including counterparty signature collection, notarisation and, for payment, the AIS and PIS calls to the Open Banking layer. Throughput is the number of successfully completed transactions in a run divided by the elapsed time of that run, where elapsed time is measured in the same way as for latency.
 
\begin{figure}
    \centering
    \includegraphics[width=1\linewidth]{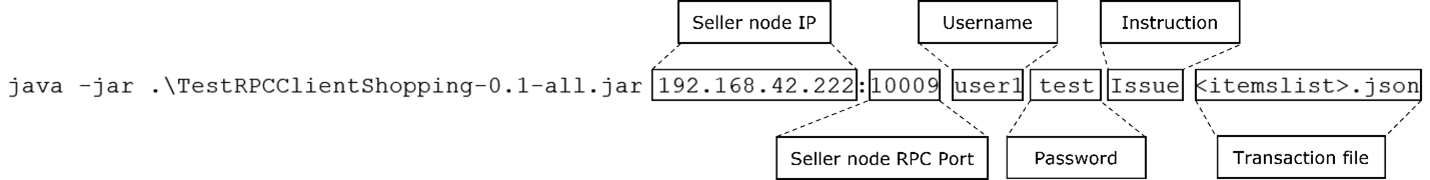}
    \caption{RPCClient command syntax}
    \label{fig:rpc}
\end{figure}

After every repetition the ledger was cleared and the nodes and accounts
recreated from scratch, so each repetition executes against an empty vault and
no state accumulates across runs. No JVM warm-up period preceded measurement. The experimental setup is summarised in Figure \ref{fig:expsetup}. The syntax of the command to spawn the clients and initiate transactions is shown in Figure \ref{fig:rpc} where:

\begin{figure}
    \centering
    \includegraphics[width=1\linewidth]{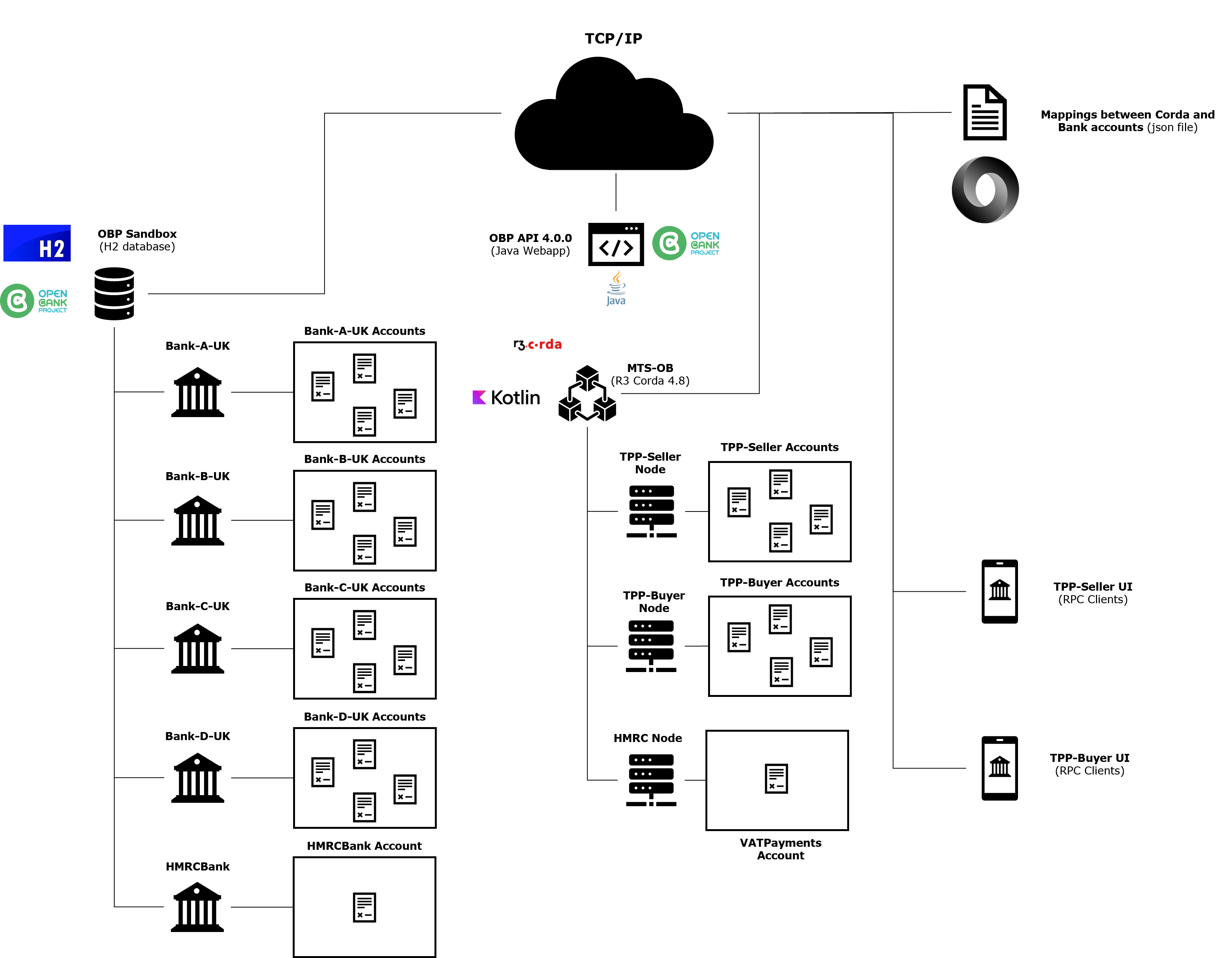}
    \caption{Experimental Setup}
    \label{fig:expsetup}
\end{figure}

\begin{itemize}
    \item Seller node IP: The IP address of the Corda node where the seller’s account belongs to.
    \item Seller node RPC port: The RPC port used by the node for intercepting incoming Remote Procedure Calls.
    \item Username: The user that authenticates to the Seller node for issuing the command.
    \item Password: The password for the user specified by Username.
    \item Instruction: The transaction that the node must generate. Can be either Issue or Pay.
    \item Transaction File: A json file with the invoice contents. See Listing \ref{lst:sampletxfile} for a sample transaction file.
\end{itemize}

\begin{lstlisting}[language=json, caption={Sample transaction file}, label={lst:sampletxfile}]
[
  {
    "amount": 10,
    "buyer": "Cassey",
    "payFrom": "HMRC",
    "shoppingList": [
      {
        "item": "Groceries",
        "price": 4,
        "quantity": 2,
        "vatRate": 20
      },
      {
        "item": "Children's Clothing",
        "price": 1,
        "quantity": 2,
        "vatRate": 0
      }
    ],
    "whoAmI": "ElectricAndGas"
  },
  {
    "amount": 13,
    "buyer": "Cassey",
    "payFrom": "HMRC",
    "shoppingList": [
      {
        "item": "Books",
        "price": 6,
        "quantity": 2,
        "vatRate": 0
      },
      {
        "item": "Books",
        "price": 1,
        "quantity": 1,
        "vatRate": 5
      }
    ],
    "whoAmI": "SmallBusiness"
  }
]

\end{lstlisting}

\subsubsection{Analysis}
 
Because the two factors interact, results are analysed at each transaction-load
level separately rather than by averaging across levels. Averaging would combine
conditions with different total offered work — at $cl = 10$ the levels span 100
to 1{,}000 invoices, at $cl = 100$ they span 1{,}000 to 10{,}000 — and would
suppress between-level variation that is itself a finding. Linear relationships
are therefore reported as the distribution of gradients and coefficients of
determination across the ten levels, obtained by ordinary least-squares
regression. Where a single figure over all 100 measurements is quoted,
transaction-load variation is retained as residual scatter rather than removed
by averaging. Full per-level results are given in Table~\ref{tab:regression} and
summarised in Table~\ref{tab:summary}.

\begin{table}[h]
\caption{Per-level regression results against client population $cl$, giving
the full fitted line $la(cl) = a + b \cdot cl$ for latency at each
transaction-load level. Latency slope $b$ is in seconds per additional
client; throughput gradient is in TPS per additional client. All latency
fits are significant at $p<0.001$.}
\label{tab:regression}
\centering\small
\begin{tabular}{r r r r r r r r}
\hline
& \multicolumn{3}{c}{Latency: $la = a + b \cdot cl$} & \multicolumn{2}{c}{Throughput} & \multicolumn{2}{c}{Throughput trend} \\
$tx$ & $a$ (s) & $b$ (s/client) & $R^2$ & mean (TPS) & $1/tr$ (s) & gradient & $R^2$ \\
\hline
\multicolumn{8}{l}{\textit{Issuance}} \\
10 & -5.98 & 0.354 & 0.972 & 5.113 & 0.196 & -0.0912 & 0.774 \\
20 & -5.90 & 0.484 & 0.989 & 3.242 & 0.308 & -0.0474 & 0.552 \\
30 & -4.46 & 0.501 & 0.995 & 2.732 & 0.366 & -0.0294 & 0.468 \\
40 & -3.15 & 0.504 & 0.996 & 2.444 & 0.409 & -0.0198 & 0.403 \\
50 & -3.06 & 0.527 & 0.996 & 2.217 & 0.451 & -0.0107 & 0.543 \\
60 & -2.36 & 0.523 & 0.997 & 2.319 & 0.431 & -0.0157 & 0.603 \\
70 & -3.65 & 0.562 & 0.998 & 2.119 & 0.472 & -0.0101 & 0.511 \\
80 & -3.05 & 0.560 & 0.996 & 1.946 & 0.514 & -0.0050 & 0.294 \\
90 & -1.48 & 0.553 & 0.998 & 1.792 & 0.558 & -0.0007 & 0.011 \\
100 & -2.25 & 0.544 & 0.998 & 1.999 & 0.500 & -0.0048 & 0.780 \\
\hline
\multicolumn{8}{l}{\textit{Payment}} \\
10 & -3.24 & 0.951 & 0.989 & 1.085 & 0.921 & -0.0028 & 0.529 \\
20 & -4.44 & 1.028 & 0.993 & 1.068 & 0.936 & -0.0032 & 0.525 \\
30 & -3.54 & 1.027 & 0.997 & 1.058 & 0.945 & -0.0023 & 0.749 \\
40 & -2.61 & 1.019 & 0.999 & 1.030 & 0.971 & -0.0017 & 0.717 \\
50 & +0.00 & 0.948 & 0.987 & 1.055 & 0.948 & -0.0015 & 0.559 \\
60 & -1.75 & 0.993 & 0.999 & 1.045 & 0.957 & -0.0010 & 0.147 \\
70 & -1.65 & 0.991 & 0.999 & 1.032 & 0.969 & -0.0021 & 0.420 \\
80 & -1.23 & 0.990 & 0.999 & 1.016 & 0.984 & -0.0013 & 0.296 \\
90 & -0.93 & 0.991 & 0.999 & 1.051 & 0.952 & -0.0014 & 0.272 \\
100 & -1.34 & 0.988 & 0.996 & 1.008 & 0.992 & -0.0020 & 0.393 \\
\hline
\end{tabular}
\end{table}
 
\begin{table}[h]
\caption{Summary of scalability results. Gradients are with respect to client
population $cl$; ranges are across the ten transaction-load levels tested.
Pooled equations are fitted over all 100 measurements per operation.}
\label{tab:summary}
\centering\small
\begin{tabular}{p{0.38\linewidth} p{0.26\linewidth} p{0.26\linewidth}}
\hline
& \textbf{Issuance} & \textbf{Payment} \\
\hline
Pooled equation ($n=100$) & $la = -3.53 + 0.511\,cl$ & $la = -2.07 + 0.992\,cl$ \\
Pooled latency $R^2$ & 0.903 & 0.993 \\
Latency gradient range & 0.354--0.562 & 0.948--1.028 \\
Gradient spread & 41\% of mean & 8\% of mean \\
Gradient vs $tx$ & $R^2=0.64$, $p=0.006$ & $R^2=0.01$, $p=0.78$ \\
Per-level latency $R^2$ & 0.972--0.998 & 0.987--0.999 \\
Latency variance from $cl$ / $tx$ & 90.5\% / 8.0\% & 99.4\% / 0.2\% \\
\hline
Peak throughput (TPS) & 11.30 & 1.42 \\
Saturation throughput (TPS) & 2.15 $\pm$ 0.47 ($cl\geq40$) & 1.045 $\pm$ 0.090 (full range) \\
Throughput linear in $cl$? & no ($R^2=0.19$) & no ($R^2=0.39$) \\
Throughput variance from $cl$ / $tx$ & 29.4\% / 36.1\% & 57.2\% / 6.4\% \\
Mean deviation from $cl$ & 8.2\% & 5.4\% \\
Gradient vs $1/tr$ & $R^2=0.87$ & invariant \\
Transactions rejected & 0 & 0 \\
\hline
\end{tabular}
\end{table}

\section{Results}
This section reports the performance of the MTS-OB prototype under increasing
workload, and establishes whether the cost of servicing additional load grows
predictably in the sense set out in Section~\ref{sec:expdesign}.

\subsection{Issuing an Invoice via MTS-OB}\label{issueexp}

\begin{figure}
    \centering
    \includegraphics[width=1\linewidth]{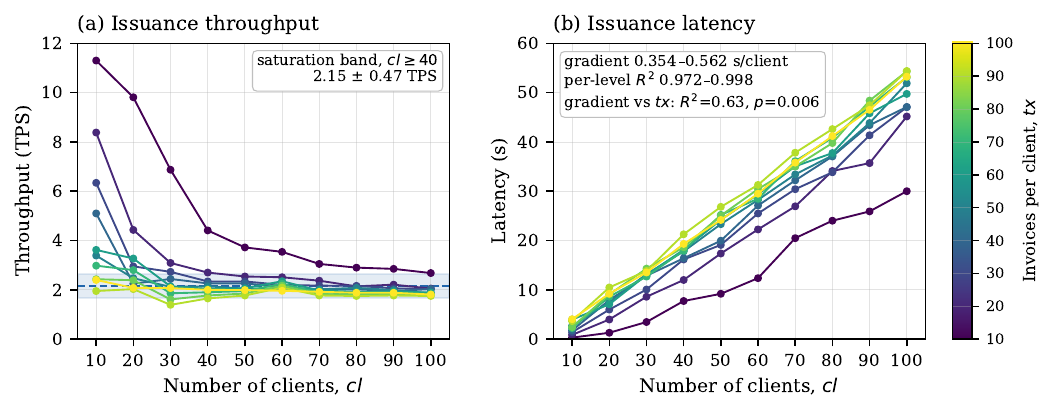}
    \caption{Issuance performance under increasing client population. Each curve is one transaction-load level, coloured from $tx=10$ (dark) to $tx=100$ (light). (a) Throughput, converging beyond $cl=40$ into a saturation band of
$2.15 \pm 0.47$\,TPS (shaded). (b) Latency, linear at every load level ($R^2$ 0.972--0.998) with a gradient increasing systematically with transaction load.}
\label{fig:issueresults}
\end{figure}

Throughput and latency for invoice issuance are shown in
Figure~\ref{fig:issueresults}, with each curve corresponding to one
transaction-load level.

Latency increases linearly with client population at every level tested. The
pooled fit over all 100 measurements is
\begin{equation}
    la(cl) = -3.53 + 0.511\,cl, \qquad R^2 = 0.903,
\end{equation}
and per-level fits give $R^2$ between $0.972$ and $0.998$, all significant at
$p < 0.001$. The negative intercept is an artefact of fitting over the tested
range $cl \in [10,100]$ rather than a physically meaningful latency at zero
clients, and should not be extrapolated below $cl = 10$. The gradient is not
constant across levels: it rises from $0.354$\,s per client at $tx = 10$ to
$0.562$\,s per client at $tx = 70$, and the increase is systematic ($R^2 = 0.64$, $p = 0.006$ for gradient against $tx$). Each additional endpoint adds a fixed increment to issuance time, with no superlinear growth at the top of the range, so the client capacity meeting a target latency can be estimated once the operating load level is fixed.
 
Throughput does not correlate with the client population ($R^2 = 0.19$ over all
measurements) and instead saturates. Peak throughput is $11.30$\,TPS at
$tx = 10$, $cl = 10$. Throughput falls steeply up to $cl = 40$, beyond which all
ten curves converge into a band of $2.15 \pm 0.47$\,TPS sustained to the top of
the range. At $cl = 10$ the levels are widely separated, spanning $1.95$ to $11.30$\,TPS, whereas beyond $cl = 40$ the load level has not effect on performance. This happens because the node operates at capacity, and the arrival rate exceeds the rate at which transactions can be validated, signed and notarised. Subsequent transaction requests are queued before they are processed. The same profile has been reported for Hyperledger Fabric \cite{ENAREABANG2024101217}. Importantly no transaction was rejected at any factor level and no run failed to complete. Unlike latency, issuance throughput is influenced more by transaction load than by client population, with $36.1\%$ of variance attributable to $tx$ against
$29.4\%$ to $cl$. 

\subsection{Paying an Invoice via MTS-OB}\label{payexp}

\begin{figure}
    \centering
    \includegraphics[width=1\linewidth]{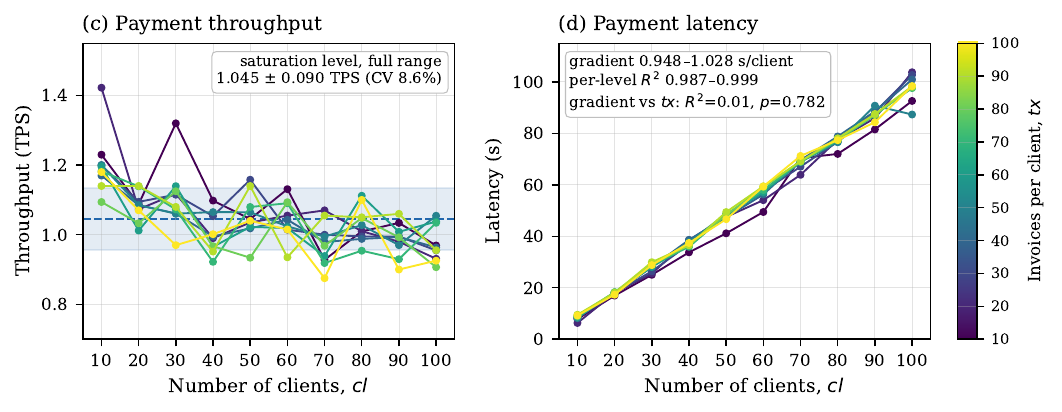}
    \caption{Payment performance under increasing client population, coloured as in Figure~\ref{fig:issueresults}. (c) Throughput, saturated across the full range at $1.045 \pm 0.090$\,TPS (shaded). (d) Latency, linear at every load level ($R^2$ 0.987--0.999) with an invariant gradient of approximately one second per client.}
    \label{fig:payresults}
\end{figure}

Throughput and latency for invoice payment are shown in
Figure~\ref{fig:payresults}. Payment performance is lower than issuance
performance throughout. Two mechanisms account for the difference. First, payment incurs extra computational overhead for the additional calculations, necessary checks, and for updating the balances in Corda accounts. Second,
payment requires two synchronous Open Banking calls — an AIS balance query and a
PIS transfer request — together with the corresponding updates to the OBP
database, none of which is incurred during issuance. The additional cost is
attributable to the split-payment mechanism itself and to external API latency,
rather than to any difference in ledger behaviour.
 
Payment latency increases linearly with client population at every
transaction-load level. The pooled fit over all 100 measurements is
\begin{equation}
    la(cl) = -2.07 + 0.992\,cl, \qquad R^2 = 0.993,
\end{equation}
and per-level $R^2$ ranges between $0.987$ and $0.999$, all significant at
$p < 10^{-8}$. As with issuance, the negative intercept reflects fitting over the tested range $cl \in [10,100]$. In contrast to issuance, the gradient is essentially
invariant: it varies only between $0.948$ and $1.028$\,s per client, with no systematic relationship to transaction load ($R^2 = 0.01$, $p = 0.78$). The cost of an additional client on the payment path is
therefore approximately one second, regardless of how many invoices that client
submits.
 
Payment throughput does not decline appreciably with load. It remains at
$1.045 \pm 0.090$\,TPS (CV $8.6\%$) across the full range, with individual
measurements between $0.875$ and $1.423$\,TPS. A weak negative trend with client
population is detectable but accounts for little of the variation
($R^2 = 0.39$), amounting to roughly $17\%$ decline across a tenfold increase in
client population, and only six of the ten per-level regressions reach
significance. Transaction load has no measurable effect ($R^2 = 0.04$). The
payment path is therefore already saturated at the lowest client population
tested and sustains its rate thereafter.

\subsection{Threats to validity}\label{sec:validity}

Five limitations bound the interpretation of these results. 

\begin{enumerate}
\item The prototype uses Corda Open Source 4.8, which runs a single flow worker per node. Corda Enterprise permits multiple flow workers, and R3 reports
approximately tenfold throughput scaling with core count on that basis, while noting that per-flow execution time is unchanged between editions. The saturation levels reported here are therefore a property of the flow-concurrency configuration rather than of the MTS-OB design or of the Corda platform generally. The testbed also persists transactions to in-process H2 databases, which is the DBMS employed in Corda Open Source 4.8. According to R3: "The H2 database [...] cannot handle the throughput that a serious performance test will generate on any sensible hardware and [...] will throttle the throughput of the node. To test actual performance of a Corda system, it is required to use an enterprise level database." \cite{r3_corda_perf_tuning_2026}.

\item The notary is a validating notary, which re-executes contract verification over
the transaction and its dependency chain before notarising. This is
substantially more expensive than non-validating notarisation and contributes to
the saturation levels reported; it was selected for the assurance it provides
rather than for performance.

\item Absolute figures reflect a single machine hosting the RPC clients, four Corda
nodes and the OBP sandbox concurrently. The measurements characterise one node
servicing a growing client population and do not establish capacity gained by
adding nodes. In deployment each TPP operates its own node and serves its own
clients, so aggregate capacity is not bounded by the single-node figures
reported here. Establishing that figure requires a distributed testbed and is
identified as future work.

\item The Open Banking layer is an OBP-API sandbox rather than a production bank
interface. The sandbox implements the OBL-specified endpoints in accordance with the Open Banking Standard \cite{obie2024standards} but its response-time is not
representative of a production deployment, and the payment-side latencies
reported here should be read as a lower bound. 

\item Finally, no JVM warm-up period preceded measurement \cite{Lion2016}, so just-in-time compilation may not have reached steady state in the shortest runs. This is a plausible contributor to the $tx = 10$ issuance results, which depart from the pattern
established at every other load level with a latency gradient of $0.354$\,s per
client against $0.48$–$0.56$ elsewhere. Dispersion figures quoted throughout
describe variation across transaction-load levels at fixed client population;
run-to-run variability across the ten repetitions is not reported.
\end{enumerate}

\section{Discussion}
The majority of research in payments and money innovation is directed towards applications of DLT to build alternative forms of money such as cryptocurrencies, stablecoins and CBDC's. This development has largely been incentivised by the consequences of the 2008 financial crisis owed to the inability of governments and central banks to control the risk contagion with the policy tools available at the time. The literature review presented herein shows that the proposed DLT-based alternative forms of money are unsuitable candidates to support innovation in either the payment or the tax systems in place. On the other hand, the financial crisis was followed by an influx of new regulations in e-invoicing and open banking and standards such as EN16931 and ISO20022. These have helped shape the economic and technological environments required for supporting transformative innovation of the money system by mobilising trade and payments data. Moreover, when the Peppol network becomes mandatory, invoices will be available in the EN16931 standard format, streamlining data field extraction for automated processing, such as VAT calculation, as presented in this paper. 

It follows that the reinvention of money in the form of cryptocurrencies and stablecoins is unnecessary. The technology and regulations in place are sufficient to transform money in a way which brings benefits for all stakeholders in the economy without causing additional risks or compromises to the resilience of the financial system. The key is to capitalise on the innovative aspects of DLT, the transparency and interoperability of the hitherto siloed core banking system which is permitted by new payment regulations and standards.

The aim of this paper is to design, build and evaluate a permissioned distributed ledger architecture which demonstrates the feasibility of extending the functionality of money to support automated tax payments using bank deposit money. Building upon previous research where it was demonstrated that automated tax payments are feasible with the R3 Corda DLT \cite{louvieris2024making}, we have augmented the original MTS PoC with payments using real money from bank deposits, using certified Open Banking APIs.

The realisation of real-time VAT compliance via split payments necessitates a fundamental shift in how money is defined within a digital economy. While the traditional functions of money—medium of exchange, unit of account, and store of value—have not changed since antiquity, the MTS-OB architecture demonstrates that money must now also act as a real-time information sensor. By integrating tax functionality directly into the payment rail, the system leverages the "informational value" of transaction metadata, such as ISO 20022 structured fields and e-invoices to automate compliance. 

Payment performance is lower than issuance performance throughout the experiments. This is due to the mandatory triple-ledger synchronisation between the Buyer, Seller, and HMRC nodes, as well as the external latency introduced by Open Banking API communication and database updates. However, the behaviour observed in both metrics is a vital finding where latency rises linearly with client population at every load level, and throughput saturates rather than collapses. The architecture is therefore predictable and tractable. Performance limits are not symptoms of architectural failure because all flows are executed on a single-node causing resource contention within the testbed. This would not occur in a distributed production environment or with multiple flow workers.

\section{Conclusions}
This research demonstrates the technical and regulatory feasibility of integrating real-time VAT compliance into the existing UK payment ecosystem offering a path towards automated tax compliance and a potential reduction in the VAT gap without compromising financial stability. The MTS-OB architecture is the first-of-its-kind infrastructure capability which integrates the unique permission and access–control management facilitated by Distributed Ledger Technology with the Open Banking layer to successfully transform bank deposit money—which constitutes 96\% of the current money supply—into a sensor for economic intelligence without disrupting core banking operations or requiring new legislation. Instead of volatile and unregulated cryptoassets, this reform maintains financial stability while providing the "hard to get tax wrong" infrastructure necessary for modernizing tax administration.

The scalability study confirms that the architecture can maintain operational stability under increasing transaction loads, with latency that scales predictably and throughput that saturates without collapsing. Ultimately, this framework offers a path toward closing the £12.1 billion VAT gap by replacing asynchronous, error-prone reporting with a synchronous, "invisible" compliance layer built directly into the fabric of the money system. This confirms MTS-OB offers the real prospect of unobtrusively building tax functionality into money (i.e. a medium of real-time taxation) without causing financial instability.

\section{Future Work}
Positioning the MTS-OB architecture as an invisible layer between payment rails, makes its application possible on many other tax purposes outside VAT. The PAYE system calculates and deducts income taxes from salaries which are payable every month; the MTS-OB can accommodate split PAYE payments by adopting a new CorDapp, which uses the appropriate calculation mechanism for income taxes \cite{govuk_paye_employers}. Also in the UK, the Carbon Border Adjustment Mechanism which will apply from 2027, which means that UK traders may be required to pay taxes for carbon intensive goods they import from abroad \cite{govuk_cbam_intro}. A new CorDapp deployed on the MTS-OB architecture can employ the mechanisms for calculating carbon taxes and triggering split payments via Open Banking, when transactions involving carbon intensive goods are settled via the UK's payment rails. Similar withholdings of taxes that could be serviced with MTS-OB apply under the Construction Industry Scheme, where contractors deduct money for tax and National Insurance purposes from subcontractors' payments and pass it to the HMRC \cite{govuk_cis_overview}.

\section{Acknowledgements}
The authors would like to thank the HMRC Innovation Team, Chief Technology \& Design Office - CDIO for their guidance and for facilitating project workshops and meetings.

\bibliographystyle{elsarticle-num}
\bibliography{references}
\end{document}